\documentclass[entropy,article,accept,moreauthors,pdfm,10pt,a4paper]{mdpi} 
\firstpage{1} 
\lastpage{\pageref*{LastPage}}
\articlenumber{20}
\doinum{10.3390/e18010020}
\pubvolume{18}
\pubyear{2016}
\externaleditor{Academic Editor: Stefano Pirandola}
\history{Received: 21 May 2015; Accepted: 30 December 2015; Published: date}

\usepackage{graphicx,epsf,amsmath,amssymb,times,epsf,epsfig}
\usepackage{soul}

\usepackage{graphicx}
\usepackage{caption}
\usepackage{subcaption}
\graphicspath{{./figures/}}

\Title{Trusted Noise in Continuous-Variable Quantum Key Distribution: A Threat and a Defense}

\Author{Vladyslav C. Usenko * and Radim Filip }

\address[1]{%
 Department of Optics, Palack\'y University, 17. listopadu 1192/12, 77146 Olomouc, Czech Republic; \linebreak filip@optics.upol.cz }

\corres{Correspondence: usenko@optics.upol.cz; Tel.: +420-585-634-248}

\abstract{We address the role of the phase-insensitive trusted preparation and detection noise in the security of a continuous-variable quantum key distribution, considering the Gaussian protocols on the basis of coherent and squeezed states and studying them in the conditions of Gaussian lossy and noisy channels. The influence of such a noise on the security of Gaussian quantum cryptography can be crucial, even despite the fact that a noise is trusted, due to a strongly nonlinear behavior of the quantum entropies involved in the security analysis. We recapitulate the known effect of the preparation noise in both direct and reverse-reconciliation protocols, as well as the detection noise in the reverse-reconciliation scenario. As a new result, we show the negative role of the trusted detection noise in the direct-reconciliation scheme. We also describe the role of the trusted preparation or detection noise added at the reference side of the protocols in improving the robustness of the protocols to the channel noise, confirming the positive effect for the coherent-state reverse-reconciliation protocol. Finally, we address the combined effect of trusted noise added both in the source and the detector.}


\begin{document}


\section{Introduction}

Quantum key distribution (QKD; see \cite{Gisin2002,Scarani2009} for reviews) is the branch of quantum information science, whose goal is to develop methods (protocols) that allow two trusted parties to share a random secret key so that its security is provided by the very laws of quantum physics. The key can be then used in the one-time pad cryptographic system \cite{Vernam1926}, which was shown to be information-theoretically secure \cite{Shannon1949}. Therefore, QKD provides a quantum-cryptographic physics-based solution as an alternative to the classical asymmetrical cryptosystems, which are currently widely used, but are based on the mathematical complexity assumptions.

The first theoretical idea of QKD was the BB84 protocol (named after its authors Bennett and Brassard)~\cite{Bennett1984} based on the use of the discrete-variable (DV) quantum systems, namely single photons, so that the key bits were encoded to (and obtained from) the measurements of the polarization degree of freedom in the two randomly-switched bases. BB84 and its modifications, such as B92 (named after its author Bennett)~\cite{Bennett1992}  or SARG  (named after its authors Scarani, Ac\'in, Ribordy and Gisin)~\cite{Scarani2004}, was used as the basis of DV QKD in the {prepare-and-measure} (P~\& M) implementations typically performed with weak coherent pulses or heralded singe photon sources using, besides polarization, e.g., phase or time encoding \cite{Bennett1992,Brendel1999}. The presence of the additional photons in the weak coherent pulses was shown to be a potential vulnerability of DV QKD in the case of photon number splitting attacks, and the use of the decoy states was suggested to overcome such a threat \cite{Lo2005}. On the other hand, the use of photonic entanglement was suggested in the {entanglement-based} (also called {EPR-based} after the famous Einstein--Podolsky--Rosen paradox \cite{Einstein1935}) E91 protocol \cite{Ekert1991}, where trusted parties are supposed to perform polarization measurements in the certain sets of bases on the two spatially distant non-classically correlated photons. By verifying a loophole-free violation of the Bell inequalities \cite{Bell1964,Clauser1969} from the measurements in one set of bases, the trusted parties can ensure the security of the key obtained from the measurements in another one. Although the E91 protocol can be seen as an alternative way of implementing BB84 \cite{Bennett1992a}, it also offers potential device independence based on the quantum nonlocality. While the security of the protocols was first considered against particular eavesdropping strategies, it was later extended on the general information-theoretical security proofs against collective \cite{Devetak2005} and coherent attacks \cite{Kraus2005} and then extended to composable security (being the systematic way of specifying the security requirements of cryptographic tasks)~\cite{Muller2009} for certain DV QKD protocols \cite{Renner2007,Christandl2009,Tomamichel2012}.

As an alternative to the DV coding, the use of the continuous variables (CVs; see \cite{Braunstein2005} for a review) of multiphoton Gaussian states of light (see \cite{Weedbrook2012a} for a review on Gaussian quantum information) was suggested in the CV QKD protocols. Typically the continuous quadrature observables are used to encode key bits, and subsequently, the {homodyne detection} of the quadrature (or {heterodyne} measurement of two complementary quadratures simultaneously) is applied contrary to the photon-counting measurement in DV QKD. Alternatively, polarization \cite{Lorenz2004} or photon-number coding \cite{Funk2002,Zhang2003,Usenko2005,Usenko2007,Usenko2010} can in principle be used. The preliminary version of CV QKD was firstly suggested \cite{Ralph1999} and studied against particular eavesdropping strategies \cite{Ralph2000} by Ralph based on the binary quadrature displacement of coherent state produced by a laser or two-mode quadrature-squeezed states, which can be produced by an optical parametric oscillator. In the latter case, both the entangled modes were supposed to be sent to the remote trusted party for a homodyne measurement. Later, a protocol based on the quadrature modulation of a single-mode squeezed states was suggested by Hillery \cite{Hillery2000} and studied against arbitrary attacks taking into account error correction by Gottesman and Preskill \cite{Gottesman2001}, while the encoding of a pre-determined key using entangled states and verification of nonclassical correlations similarly to E91 protocol to prove security against particular eavesdropping attacks was suggested by Reid \cite{Reid2000}. The CV QKD protocol based on the two-mode entangled beams shared between the trusted parties who use amplitude or phase measurements to decode binary data was alternatively suggested by Silberhorn \emph{et al.} \cite{Silberhorn2002}. The security of the protocols was shown against certain eavesdropping strategies mainly in purely attenuating channels and typically relied on the uncertainty principle, which does not allow a potential eavesdropper to measure both quadratures precisely and simultaneously. The CV states of light were also suggested by Cerf \emph{et al.} \cite{Cerf2001} to distribute a continuous Gaussian key with security being shown against the whole class of individual attacks using the optimal entangling cloner attack. Importantly, it was then shown by Grosshans and Grangier \cite{Grosshans2002} that Gaussian protocols secure against individual attacks can be implemented even with coherent states with no need in nonclassicality, such as squeezing. The Gaussian protocols were however limited by the 50\% loss in the optical link, because it otherwise led to the information advantage of an eavesdropper with respect to the sender trusted party when the standard {direct reconciliation} (DR) of data was used, \emph{i.e.}, when the remote trusted party was correcting its errors to exactly reproduce the dataset at the trusted sender side. The problem can be solved by using post-selection, as was shown by Silberhorn \emph{et al.} \cite{Silberhorn2002a}. Remarkably, the {reverse reconciliation} (RR) can be used and allows achieving theoretical security of coherent-state CV QKD upon any channel loss, as shown by Grosshans \emph{et al.} \cite{Grosshans2003}. The RR protocol was extended by Weedbrook \emph{et al.}  \cite{Weedbrook2004} to the heterodyne detection allowing coherent-state CV QKD with no bases switching. Recently, the simplified unidimensional CV QKD protocol based on a single-quadrature modulation was suggested by Usenko and Grosshans \cite{Usenko2015}. The~Gaussian individual attacks were shown to be optimal for the Gaussian CV QKD protocols by Grosshans and Cerf~\cite{Grosshans2004}. It~was then an important step in CV QKD when the security of the Gaussian protocols was generalized and proven against the Gaussian collective attacks by Navascu\' es \emph{et al.} \cite{Navascues2006} and by Garc\' ia-Patr\' on and Cerf \cite{Garcia2006}. Such~attacks were shown optimal using the extremality of Gaussian states~\cite{Wolf2006}, studied by Pirandola \emph{et al.} \cite{Pirandola2008,Pirandola2009}, and then extended to the general attacks using the de Finetti theorem by Renner and Cirac \cite{Renner2009}. This established the clear framework for theoretical security analysis of the CV QKD protocols in asymptotic regime, but many issues related to the real implementation of the protocols remained open. Differently from many other CV quantum information protocols, such as CV quantum teleportation \cite{Braunstein1998} in the Gaussian regime \cite{Chizhov2002,Fiurasek2002,Pirandola2006} or Gaussian CV quantum cloning \cite{Cerf2000,Braunstein2001,Fiurasek2001}, the Gaussian CV QKD involves more complex aspects of the Gaussian quantum states and measurements due to the strong nonlinearity of the quantum entropies, involved in the analysis of the security of the protocols. 

The information-theoretical research in CV QKD is currently focused on extending the security proofs against general attacks on the finite-size regime, which is always the case in real QKD systems. The first step in this direction was done by Leverrier \emph{et al.} \cite{Leverrier2010}, who considered finite-size effects in coherent-state CV QKD secure against collective attacks mainly focusing on the channel estimation. The channel estimation was recently optimized for CV QKD with arbitrary signal states and Gaussian modulation by Ruppert~\emph{et~al.}~\cite{Ruppert2014}. The optimality of Gaussian collective attacks against the Gaussian CV QKD was confirmed in the finite-size regime by Leverrier and Grangier \cite{Leverrier2010a}. The security of CV QKD against arbitrary attacks was shown using Gaussian post-selection (incorporated to the protocol) in the asymptotic limit by Walk \emph{et al.} \cite{Walk2013} using the equivalence between post-selection and noiseless linear amplification, also shown by Fiura\v sek and Cerf \cite{Fiurasek2012}.
The composable security of the protocol based on the two-mode squeezed vacuum state against arbitrary attacks in the finite-size regime was shown by Furrer \emph{et al.} \cite{Furrer2012} using entropic uncertainty relations for smooth entropies. A similar proof was later constructed by Furrer for squeezed-state protocol \cite{Furrer2014}, preceded by the security proof of coherent-state protocol against arbitrary attacks in the finite-size regime derived by Leverrier~\emph{et~al.} using post-selection and phase-space symmetries \cite{Leverrier2013}. The composable security proof for the coherent-state protocol against collective attacks (and general attacks using post-selection or the de Finetti theorem) was also recently derived by Leverrier \cite{Leverrier2015}. The general security bounds derived up to now are however fragile against imperfections, apply to particular cases of coherent or squeezed signal states and often require additional procedures, such a post-selection or signal monitoring. Hence, the study of the effective general composable security proofs for CV QKD protocols in the finite-size regime is still ongoing.

Another line of information-theoretical research concerned with CV QKD protocols is dedicated to the post-processing algorithms, particularly error correction codes. Error correction is a demanding procedure for Gaussian random variables, especially in the regime of low signal-to-noise ratio (SNR), RR and one-way classical communication. Realistic error correction in this regime scales down the mutual information between the trusted parties and therefore limits the secure key rate, which was first pointed out for CV QKD protocols by Heid and L\"utkenhaus \cite{Heid2006}. It appeared to be one of the main limiting factors in the early practical implementations of CV QKD, as shown in \cite{Lodewyck2007} by Lodewyck \emph{et al.}, where the 10\% reduction of mutual information due to imperfect error correction along with other imperfections resulted in the secure distance of the coherent-state protocol being limited by 25 km, similarly to another field test of the coherent-state CV QKD prototype performed by Fossier \emph{et al.} \cite{Fossier2009}. The novel codes for reconciliation of the Gaussian data in CV QKD were developed afterwards, such as multidimensional codes by Leverrier \emph{et al.} \cite{Leverrier2008}, polar and low density parity check codes by Jouguet~\emph{et~al.}~\cite{Jouguet2011,Jouguet2014}. By applying the optimal codes at low SNR regime, Jouguet~\emph{et~al.}~\cite{Jouguet2013} proved that the achievable distance of the coherent-state CV QKD protocols can be extended to 80~km. An alternative way to using the advanced post-processing codes can also be the state engineering. It~was in particular shown by Usenko and Filip that the CV QKD protocol with feasible squeezed states and limited Gaussian modulation can be robust against imperfect post-processing \cite{Usenko2011}. 

From the point of view of quantum-theoretical analysis, the implementation of CV QKD is naturally limited by physical effects causing the practical imperfections, such as losses and noise in the quantum communication channel. In all of the CV QKD security analysis, the channel is assumed to be fully controlled by an eavesdropper, who can purify the channel noise and compensate the channel loss; therefore, the channel is {untrusted}. As was mentioned, any level of channel loss is in principle tolerable by a perfect CV QKD protocol with RR~\cite{Grosshans2003} even for collective attacks as shown by Grosshans \cite{Grosshans2005} contrary to the DR schemes, which are limited by 50\% of channel transmittance. The presence of loss however increases the vulnerability of the protocols to other imperfections, first of all to the channel excess noise. This was theoretically demonstrated already for the case of individual attacks by Grosshans \emph{et al.} \cite{Grosshans2003a} using the equivalent entanglement-based representation of the CV QKD protocols. It was shown that the tolerable channel noise for Gaussian CV QKD protocols is upper bounded by one shot-noise unit (SNU) being the level of the vacuum fluctuations (used as a unit to characterize the amounts of quantum noise), which is more strict than the bound on the Gaussian entanglement breaking, being two SNU of excess noise. The more tight security bounds on the Gaussian channel noise (which complies with the optimality of Gaussian attacks summarized by Leverrier and Grangier \cite{Leverrier2010a}) were obtained for the case of collective attacks for coherent- and squeezed-state protocols by Navascu\'es and Ac\'in~\cite{Navascues2005} and for coherent-state protocol, including post-selection, by Heid and L\"utkenhaus \cite{Heid2007}. It was also shown by Blandino~\emph{et~al.} that the robustness of the coherent-state CV QKD protocol to the channel imperfections can be improved by the use of noiseless amplifiers \cite{Blandino2012}. 

However, channel imperfections are not the only threat to the security of the protocols coming from the physical effects in the set-up, and device imperfections may also contribute to the information on the key, which is accessible to a potential eavesdropper. There are two main approaches to studying the security of QKD with imperfect devices. One is the device-independent (DI) approach, when the protocols are designed in such a way that device imperfections have no impact on the security of the key. This is achieved by verifying the fragile nonclassical properties of the quantum states in the DI QKD, which was inspired by the above-mentioned E91 protocol \cite{Ekert1991}. The idea of using nonclassical correlations to verify the security of QKD when the devices (including the source of the quantum states) are not trusted was first stated by Mayers and Yao \cite{Mayers1998} and developed by Barrett \emph{et al.} \cite{Barrett2005} and by Ac\'in \emph{et al.} \cite{Acin2006} to prove security against no-signaling post-quantum eavesdropper, which goes beyond the typical assumption used in QKD that an eavesdropper is limited by the laws of quantum physics. The security of DI QKD protocols against collective attacks in its connection with Bell-type inequality violation was shown by Ac\'in \emph{et al.} \cite{Acin2007}, and a practical proposal based on heralded qubit amplifier for the implementation of DI QKD protocols was suggested by Gisin \emph{et al.} \cite{Gisin2010}. The full security of DI QKD was recently shown by Vazirani and Vidick \cite{Vazirani2014}. For the Gaussian CV QKD, however, the possibility to build the fully DI schemes based on the Bell-inequality violation is limited and requires non-Gaussian resources or measurements \cite{Grangier1988,Banaszek1998,Banaszek1999}. DI CV QKD protocol was recently suggested by Marshall and Weedbrook \cite{Marshall2014} based on the qubit encoding \cite{Gottesman2001a} using squeezed states and homodyne detection. The potentially more feasible measurement device-independent (MDI) QKD being a way to at least isolate detectors and prevent side-channel attacks was suggested by Braunstein and Pirandola \cite{Braunstein2012} and by Lo \emph{et al.} \cite{Lo2012} and supposes the use of an untrusted relay between the trusted parties. Differently from DI QKD, the MDI QKD can be potentially realized only with Gaussian states and measurements. Therefore, the concept of MDI QKD was recently applied to the CV protocols by Pirandola \emph{et al.} \cite{Pirandola2014arxiv,Pirandola2015} and studied by Zhang \emph{et al.} \cite{Zhang2014} and by Li \emph{et al.} \cite{Li2014}.

Contrary to DI and MDI QKD, a more practical and robust way to provide security with imperfect devices is to perform a comprehensive characterization of the devices and suggest the feasible methods allowing one to compensate or reduce the possibility of eavesdropping through the device imperfections. In this approach, alternative to DI and MDI QKD, the devices can be assumed {trusted}, which can be referred to as the device-dependent (DD) approach in QKD. In this case, the devices must be properly studied, modeled and calibrated in order to distinguish their imperfections from the influence of the untrusted quantum channel (otherwise, the so-called `paranoid' approach in DD QKD, when all of the imperfections are attributed to the channel, has typically very limited applicability, since the security of conventional DD QKD protocols in this case is quickly degraded by the imperfect devices). The {trusted noise} despite being directly inaccessible to an eavesdropper may, however, still contribute to the information leakage and limit the security of the protocols. On the other hand, the proper amounts of the trusted noise may improve the security of DD QKD protocols by decoupling an eavesdropper from the reference side of the data reconciliation. The two approaches to security of QKD with imperfect devices were recently generalized by Pirandola \cite{Pirandola2014} in their connection to quantum discord (being sufficient for DD QKD and upper bounding the secure key rate) and entanglement (being necessary for DI QKD). 

In the current paper, we review the main known results of the studies of trusted noise in CV QKD and also provide some new results filling the gaps that existed in the comprehensive analysis of the role of trusted preparation and detection noise in CV QKD. We show the negative effect of trusted noise on the receiver side of the protocols and the potentially positive role of such noise on the reference side. The novel results of the paper include: the negative role of the trusted detection noise in the DR scheme and the analytical bound on such a noise; the advantage of the squeezed-state protocol above the coherent-state one in the DR regime and the imperfect post-processing; the positive role of the preparation noise in the squeezed-state DR protocol; analysis of the joint influence of the two types of noise simultaneously present in the protocols. 

\section{Gaussian Continuous-Variable Quantum Key Distribution Protocols}

In our work we deal with the Gaussian states of the modes of electromagnetic radiation defined as the states with a Gaussian Wigner function (see \cite{Weedbrook2012a} for a review). In particular, we consider the coherent states, each being the eigenstate $a|\alpha\rangle=\alpha|\alpha\rangle$ of the annihilation operator of a given mode, as the possible signal states for CV QKD protocols. We study the one-way Gaussian CV QKD protocols based on the Gaussian modulation of the amplitude and phase quadratures, which can be defined as the real and the imaginary parts of the complex amplitude of an {\it n}-th mode: $x_n=a_n^\dag+a_n$ and $p_n=i(a_n^\dag-a_n)$ with the commutation relation $[x_n,p_n]=2i$. In~such notation, the quadrature fluctuations of a coherent (or of the vacuum) state are equal to one, which is defined as a shot noise unit (SNU). We also consider squeezed states with the fluctuations of one of the quadratures being suppressed below the shot noise. 

\begin{figure}[H]
\centering
\begin{subfigure}[b]{0.49\textwidth}
\centering
\includegraphics[width=0.85\textwidth]{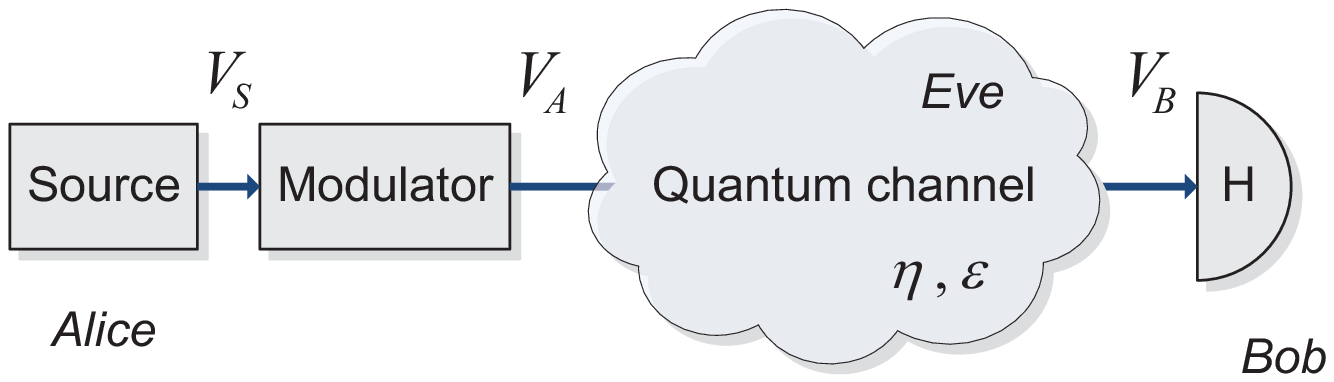}	
\caption{}
\end{subfigure}
\begin{subfigure}[b]{0.49\textwidth}
\centering
\includegraphics[width=0.85\textwidth]{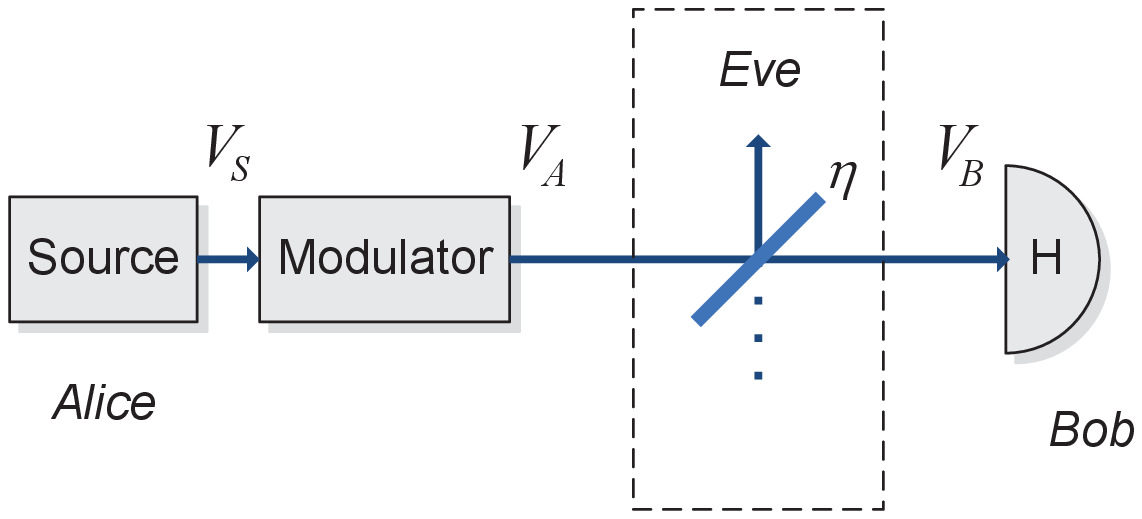}
\caption{}
\end{subfigure}
\caption{({\bf a}) Continuous variable (CV) QKD {prepare-and-measure} (P~\& M) scheme based on the signal state preparation (with variance $V_S$ in the measured quadrature) in the source and the subsequent modulation using the modulator (up to variance $V_A$ in the measured quadrature) on the side of Alice, propagation through an untrusted quantum channel with loss $\eta$ and excess noise $\epsilon$, and measurement of the resulting state (with variance $V_B$ in the measured quadrature) by Bob using homodyne detector $H$; ({\bf b}) Beam splitter model of the eavesdropping attack on P \& M CV QKD protocol for the purely attenuating channel. The beam splitter with transmittance $\eta$, corresponding to the channel loss, couples the signal to a vacuum mode. The reflected mode is available to an eavesdropper, Eve, for collective~measurement.
\label{scheme_general}}
\end{figure}

We study the one-way Gaussian CV QKD protocols as depicted in Figure \ref{scheme_general}a. The sender trusted party, Alice, prepares a signal state using a source, which can be a laser (in the case of coherent states) or an optical parametric oscillator (in the case of squeezed states). The signal states are characterized by the quadrature values $x_S,p_S$ randomly distributed around zero according to Gaussian distributions with variances $Var(x_S)=\langle x_s^2\rangle \equiv V_{xS}$ and $Var(p_S) = \langle p_s^2\rangle \equiv V_{pS}$, while coherent states saturate the uncertainty principle with $V_{xS}=V_{pS}=1$ SNU. Alice uses the amplitude and phase modulator to apply random quadrature displacements $x_M,p_M$ of amplitude and phase quadratures of the signal states, respectively, so that the resulting modulated state is characterized by quadratures $\{x,p\}_A=\{x,p\}_S+\{x,p\}_M$ with variances $Var(\{x,p\}_A)=V_{\{x,p\}S}+V_{\{x,p\}M}$, where $V_{\{x,p\}M}$ is the variance of the Gaussian modulation applied to $x$ and $p$ quadratures, respectively. All together, the state generation and modulation can be defined as the {state preparation}.

The modulated state then travels through the untrusted quantum channel, which is generally lossy and noisy, and is measured by the remote trusted party (Bob) using homodyne measurements.

Following the optimality of Gaussian collective attacks on the Gaussian CV QKD, which we discuss later, we assume a Gaussian channel, parametrized by transmittance $\eta$ and by the variance of the excess noise with respect to the input of the channel $\epsilon$. Then, the quadrature values at the output of the channel, which are measured by the remote trusted party Bob, are $\{x,p\}_B=\sqrt{\eta}(\{x,p\}_A+\{x,p\}_N)+\sqrt{1-\eta}\{x,p\}_0$, where $\{x,p\}_N$ are the quadrature values of the excess noise with variances $Var(\{x,p\}_N)=\epsilon$ (we assume the typical case of the phase-insensitive quantum Gaussian channel), and $\{x,p\}_0$ are the quadrature values of the vacuum state to which the signal is coupled in the standard model for the channel loss, $Var(\{x,p\}_0)=1$. The variances of the quadratures on the output of the channel are therefore $Var(\{x,p\}_B)=\eta[Var(\{x,p\}_A)+\epsilon]+1-\eta$. 

In the following, we assume that Bob is measuring the $x-$quadrature, which is also squeezed by Alice if the squeezed-state protocol is considered (\emph{i.e.}, the {``$x$--$x$''} protocol is implemented). We omit the discussion of the heterodyne measurement at Bob's side, because its role is mainly in adding noise to the quadrature measurement. Such noise is known to make the heterodyne protocol suboptimal in the DR case. On the other hand, the noise in the heterodyne detection can also improve the robustness of the protocols in the noisy channels in the RR case, but this improvement can be optimized beyond coupling to a vacuum, as we discuss below. Note that the trusted parties need to perform bases switching in their preparation and measurement to estimate the channel parameters in both the quadratures, but the discussion of the channel estimation is not within the scope of the current paper, so we only consider the measurements, which are contributing to the key rate, assuming that the channel parameters are properly estimated. 

The variance of the data measured by Bob in the $x-x$ protocol is then $V_B=\eta (V_A+\epsilon)+1-\eta$, where $V_A=V_S+V_M$ is the variance of that modulated at the input of the channel, $V_S$ is the signal state variance and $V_M$ is the modulation variance. The correlation between the datasets at Alice and Bob is then $C_{AB}=\sqrt{\eta}V_M$, scaled down by the channel loss. 

The security of the key in QKD protocols is based on the generalization of the Csisz\'ar--K\"orner theorem~\cite{Csiszar1978}, which states that the information shared between the trusted parties must exceed the information that a potential eavesdropper (Eve) has on the data that either of the parties possesses, then the classical algorithms can distill the secure key. It was extended to the case of quantum communication taking into account the possibility of an eavesdropper to perform the more general collective measurement (which involves storing probe states after their interaction with the signal in a quantum memory and then optimal collective measurement on the probes) by Devetak and Winter \cite{Devetak2005}, who derived the lower bound on the secure key in QKD as:
\begin{equation}
K_{DR}=\beta I_{AB}-\chi_{AE}, \\
K_{RR}=\beta I_{AB}-\chi_{BE}
\label{krgencol}
\end{equation}
where indexes $DR$ and $RR$ stand for direct and reverse reconciliation, respectively. The positivity of the lower bound Equation (\ref{krgencol}) means that the quantum protocol is secure because the classical data processing algorithms (error correction, privacy amplification) are then able to distill the secret key from the data shared between the trusted parties if they have the information advantage over an eavesdropper. 

The Gaussian states were shown to minimize the functions of the states that satisfy the continuity in the trace norm, the invariance under local unitary transformations and the strong super-additivity \cite{Wolf2006}. The lower bound on the key rate satisfies these conditions, and therefore, the attacks, which preserve the Gaussian character of the states are optimal \cite{Navascues2006,Garcia2006}. This allows us to analyze the security of CV QKD protocols considering only Gaussian channels (characterized by transmittance $\eta$ and Gaussian excess noise $\epsilon$) and the Gaussian properties of the states.

The quantity $I_{AB}$ in Equation (\ref{krgencol}) is the classical mutual information, which is the symmetrical quantity, expressed though the entropies of the input random variable $X$ and the output random variable $Y$ of a channel as $I_{XY}=H(X)+H(Y)-H(XY)=H(X)-H(X|Y)$, where $H(XY)$ is the joint entropy, quantifying the amount of transmission errors, and $H(X|Y)$ is the conditional entropy of the input with respect to the output. On the other hand, $\chi_{AE}$ and $\chi_{BE}$ are the Holevo quantities \cite{Holevo2001} in the case of DR and RR, respectively, being the capacity of a hypothetical bosonic channel between an eavesdropper and the reference side of data reconciliation. The Holevo quantity then upper bounds the classical information available to an eavesdropper upon collective attacks. Finally, the post-processing efficiency $\beta \in (0,1)$, typically taking values of up to 0.95 for the Gaussian data in RR at low SNR \cite{Jouguet2011}, is introduced as a factor to the mutual information in the lower bounds Equation (\ref{krgencol}). It defines how close the trusted parties can approach the classical mutual information, taking into account the reduction of data ensembles in the process of error correction. For the finite efficiency $\beta$, the modulation variance $V_M$ needs to be limited and optimized \cite{Lodewyck2007,Usenko2011}.

In the case of the Gaussian-distributed continuous random variables, the classical mutual information between the trusted parties can be calculated from the variances and the conditional variances of the data possessed by these parties as, for example,
\begin{equation}
I_{AB}=\frac{1}{2}\log_2{
\frac{V_A}
{V_{A|B}}
}
\end{equation}
where $V_A$ is the variance of Alice's data ($V_M$ in our case) and $V_{A|B}$ is the variance of Alice's data conditioned on the measurement results of Bob, which can be expressed through the correlation $C_{AB}$ between Alice and Bob, and the variance $V_B$ of Bob's data as $V_{A|B}=V_A-C_{AB}^2/V_B$. Taking base-two logarithms, we estimate the mutual information and other information quantities further (e.g., the lower bound on the key rate) in {bits per use of the channel}.

The Holevo bound $\chi_{YE}=S(E)-S(E|Y)$ between Eve and the reference trusted party $Y$ is calculated as the difference of the von Neumann (quantum) entropy $S(E)$ of the state available to an eavesdropper and the conditional entropy $S(E|Y)$ of the eavesdropper's state conditioned on the measurement results of the reference side of the protocol, here denoted as $Y$ (the latter entropy in the general case should be integrated over all possible outcomes of the measurement on $Y$, but collapses in the case of the Gaussian-distributed data). A generally multimode Gaussian state is explicitly described by a covariance matrix $\gamma$ with elements $\gamma_{ij}=(\langle r_i r_j \rangle+\langle r_j r_i \rangle)/2-\langle r_i\rangle\langle r_j \rangle$ containing the second moments of the quadratures in the form $r_i=\{ x_i,p_i\}$ for an {\it i}-th mode, \emph{i.e.}, variances of the modes' quadratures and quadrature correlations between the modes. Following the above-mentioned optimality of Gaussian collective attacks, it is the worst case assumption that the states involved in CV QKD are Gaussian, and therefore, the covariance matrix formalism is sufficient for security analysis of the Gaussian protocols. 

The von Neumann entropy $S(E)$ of a general $N$-mode Gaussian state can be directly calculated using the symplectic eigenvalues $\lambda_{\{1..N\}}$ of the covariance matrix $\gamma_E$ of a state through the bosonic entropic function \cite{Serafini2005} $G(x)=(x+1)\log_2{(x+1)}-x\log_2{x}$ as:
\begin{equation}
S(E)=\sum_{i=1}^N{G\Big(\frac{\lambda_i-1}{2}\Big)}
\end{equation}

Similarly, the conditional entropy is in the case of the Gaussian states straightforwardly calculated though the symplectic eigenvalues of the conditional covariance matrix $\gamma_{E|Y}$, calculated after the homodyne measurement in $x$-quadrature on mode $Y$ as:
\begin{equation}
\gamma_{E|Y}=\gamma_{Y}-\sigma_{EY}\left(X\cdot\gamma_{Y}\cdot X\right)^{MP}\sigma_{EY}^T\: \label{cvmatcond}
\end{equation}
where $\sigma_{EY}$ is the correlation matrix between the mode(s) accessible to an eavesdropper and the mode $Y$ measured by a trusted party; the diagonal matrix $X=Diag(1,0)$ (similarly, matrix $P=Diag(0,1)$ would stand for the measurement in $p$-quadrature); and $MP$ stands for the Moore--Penrose (pseudo-)inverse of a matrix~\cite{Penrose1955}, which is used because the matrix $X\cdot\gamma_{Y}\cdot X$ is singular.

The symplectic eigenvalues of the covariance matrices can be analytically found using the Williamson's form of a covariance matrix for one- and two-mode matrices \cite{Serafini2004} or numerically for a higher number of modes.

In the P \& M scheme assuming the symmetrical modulation of both the quadratures of a signal state with the diagonal covariance matrix $\gamma_S=Diag(V_S,1/V_S)$ with the same variance $V_M$, the covariance matrix of the modulation data possessed by Alice is given by $\gamma_A^{mod}=Diag(V_M,V_M)$. After the channel characterized by loss $\eta$ and excess noise $\epsilon$, the signal state measured by Bob is characterized by the matrix $\gamma_B=Diag\big(\eta[V_S+V_M+\epsilon]+1-\eta,\eta[1/V_S+V_M+\epsilon]+1-\eta\big)$. The mutual information between Alice and Bob is then given~by:
\begin{equation}
I_{AB}=\frac{1}{2}\log_2{\Bigg[1+\frac{\eta V_M}{\eta(V_S+\epsilon)+1-\eta}\Bigg]}
\end{equation}

In the simple case of a purely-attenuating channel ($\epsilon=0$), the state $\gamma_E$ is explicitly defined as the output of the beam splitter with transmittance $\eta$, which models the channel by coupling the signal to a vacuum \cite{Grosshans2005}, as shown in Figure \ref{scheme_general}b. In this case, the covariance matrix of the state available to Eve for collective measurement is given by $\gamma_E=Diag\big([1-\eta](V_S+V_M)+\eta,[1-\eta](1/V_S+V_M)+\eta\big)$, while the matrices describing the correlation between Eve and Alice and Eve and Bob, respectively, are $\sigma_{AE}=-\sqrt{1-\eta}V_M\mathbb{I}$, where $\mathbb{I}$ is the 2 $\times$ 2 unity matrix, and $\sigma_{BE}=Diag(\sqrt{\eta(1-\eta)}[1-V_S-V_M],\sqrt{\eta(1-\eta)}[1-1/V_S-V_M])$. The straightforward calculations based on the symplectic eigenvalues of covariance matrices $\gamma_E$ and $\gamma_{E|B}$ or $\gamma_{E|A}$ result in the analytical lower bounds on the key rate Equation (\ref{krgencol}), which can be simplified in the limit of arbitrarily strong modulation $V_M\to\infty$ and in the regime of perfect post-processing $\beta=1$ as:
\begin{equation}
K_{DR}^{V_M\to\infty}=\frac{1}{2}\Bigg[\log_2{\frac{\eta}{1-\eta}}-\log_2{\frac{V_S\eta+1-\eta}{V_S(1-\eta)+\eta}}\Bigg]
\label{limkrdrvs}
\end{equation}
and: 
\begin{equation}
K_{RR}^{V_M\to\infty}=\frac{1}{2}\Bigg[\log_2{\frac{1}{1-\eta}}-\log_2{\big(\eta V_S+1-\eta\big)}\Bigg]
\label{limkrrrvs}
\end{equation}
for an arbitrary signal state with variance $V_S$ or as:
\begin{equation}
K_{DR,coh}^{V_M\to\infty}=\frac{1}{2}\log_2{\frac{\eta}{1-\eta}},K_{RR,coh}^{V_M\to\infty}=\frac{1}{2}\log_2{\frac{1}{1-\eta}}
\end{equation}
for coherent states ($V_S=1$) and as:
\begin{equation}
K_{DR,sq}^{V_M\to\infty}=\log_2{\frac{\eta}{1-\eta}},K_{RR,sq}^{V_M\to\infty}=\log_2{\frac{1}{1-\eta}}
\end{equation}
for the infinitely-squeezed states ($V_S \to 0$), which gives a lower bound on the key rate twice as large compared to the infinitely-modulated coherent states \cite{Grosshans2005}. 

In the more general case, when the channel noise is present, it is assumed that Eve is able to purify the channel noise, and so, the Holevo bound can be assessed by using the purification method \cite{Garcia2006}. When Eve is holding the purification of the channel noise, the state of the system $ABE$ shared between Alice, Bob and Eve is pure, so that $S(ABE)=0$ and, from the triangle inequality \cite{Araki2002}, directly follows that $S(E)=S(AB)$. Similarly, when Alice or Bob perform the projective measurement of the respective subsystem $A$ or $B$, the state of the systems $BE$ or $AE$ is pure, and it follows that $S(E|A)=S(B|A)$ or $S(E|B)=S(A|B)$. This allows estimating the Holevo bound from the entropic characteristics of the state shared between Alice and Bob, where all of the impurity is attributed to Eve.

\begin{figure}[H]
\centering
\includegraphics[width=0.6\textwidth]{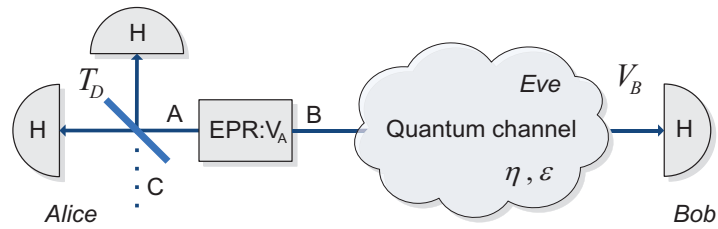}	
\caption{Entanglement-based representation of CV QKD protocols, equivalent to the P \& M scheme with the symmetrical state preparation. The entangled source $EPR:V_A$ producing two non-classically correlated modes with quadrature variances $V_A$ in modes $A$ and $B$ is placed at the sender side, Alice, who performs either homodyne measurement (corresponding to setting transmittance $T_D=1$ for a beam splitter, which couples the mode $A$ to the vacuum mode $C$), equivalent to the preparation of squeezed states, or heterodyne (corresponding to setting $T_D=1/2$), equivalent to preparation of coherent states. The rest of the scheme is the same as in the P~\&~M scenario in Figure \ref{scheme_general}a.
\label{scheme_epr}}
\end{figure}

To analyze the security in the above described case of collective attacks in a noisy channel, the equivalent EPR-based representation \cite{Grosshans2003a} is used to purify the state preparation at Alice's side, as shown in Figure~\ref{scheme_epr}. Indeed, the state of the system $AB$ must be pure before the interaction in the quantum channel with Eve's system in order to distinguish between the trusted and untrusted noise; otherwise, all of the impurity of this state would also be attributed to Eve. If the prepared state on the input of the channel is a zero-mean thermal state with variance $V_A$, being symmetric on the phase space, the state preparation is purified by a two-mode squeezed vacuum quadrature-entangled state with variance $V_A$ of the spatial modes $A$ and $B$ shared between Alice and Bob, respectively. Such a state is described by a two-mode covariance matrix of the form:
\begin{equation}
\gamma_{AB}=\left( \begin{array}{cc}
V_A\mathbb{I} & \sqrt{V_A^2-1}\sigma_z \\
\sqrt{V_A^2-1}\sigma_z & V_A\mathbb{I} 
\end{array} \right)
\label{epr}
\end{equation}
with $\sigma_z=Diag(1,-1)$.

After the homodyne measurement on mode $A$ performed by Alice and yielding the result $x_A$, the squeezed state with variance $1/V_A$ in the $x$-quadrature is conditionally prepared in mode $B$, centered around $\sqrt{1-1/V^2}(x_A,0)$ (the variance of the conditionally-prepared states in $p$-quadrature is respectively $V_A$). Therefore, the EPR-based scheme with the states Equation (\ref{epr}) and homodyne measurement at Alice becomes fully equivalent to the P \& M scheme with $V_S=1/V_A$ and $V_M=V_A-1/V_A$, \emph{i.e.}, the squeezed states in the standard protocol are modulated up to the variance of the anti-squeezed quadrature.

Similarly, if Alice performs heterodyne measurement on her mode $A$ (which is equivalent to splitting the mode $A$ on a beam splitter set to $T_D=1/2$ with another input mode $C$ being in a vacuum state, and~then measurement of both the output modes in $x$ and $p$ quadratures), resulting in the outcome $\{x_A,p_A\}$, the coherent state centered around $\sqrt{2(V-1)/(V+1)}(x_A,p_A)$ is conditionally prepared in the mode $B$. The EPR-based scheme with heterodyne detection is then fully equivalent to the P \& M scheme with coherent signal states \linebreak ($V_S=1$) and modulation variance $V_M=V_A-1$. 

Note that in the general case, the modulation variance can be independent of the signal state variance. While this is easily accessible in the case of the coherent-state protocol, where modulation variance $V_M$ is a free parameter, it is not in the standard squeezed-state symmetrical protocol where modulation depth is fixed by the antisqueezing. Therefore, the generalized EPR-based scheme, which is equivalent to the preparation of the arbitrarily-squeezed signal state $V_S \in [0,1]$ and modulation with arbitrary variance was suggested to analyze the role of squeezing in CV QKD \cite{Usenko2011} and will be used in the further analysis instead of the standard EPR-based scheme by replacing the entangled source in the cases when modulation needs to be optimized independently of the signal state squeezing.

In the standard EPR-based scheme, the entropy $S(E)=S(AB)$, being the first part of the Holevo bounds $\chi_{BE}$ and $\chi_{AE}$, is calculated from the Gaussian state of modes $AB$ described by the covariance matrix after the propagation through the channel:
\begin{equation}
\gamma_{AB}'=\left( \begin{array}{cc}
V_A\mathbb{I} & \sqrt{\eta(V_A^2-1)}\sigma_z \\
\sqrt{\eta(V_A^2-1)}\sigma_z & [\eta(V_A+\epsilon)+1-\eta]\mathbb{I} 
\end{array} \right)
\label{eprdash}
\end{equation}
(the coupling to mode $C$ being in a pure vacuum state can be omitted in this case because it does not affect the purity of the overall state shared between the trusted parties). 

Similarly, the entropy $S(E|B)=S(A|B)$, being the second part of the Holevo bound $\chi_{BE}$, can be obtained from the respective conditional matrix of mode $A$ conditioned on the measurement on mode $B$:
\begin{equation}
\gamma_{A|B}'=\left( \begin{array}{cc}
V_A-\frac{\eta(V_A^2-1)}{\eta(V_A+\epsilon)+1-\eta} & 0 \\
0 & V_A
\end{array} \right)
\end{equation}

In the case of a squeezed-state protocol, the entropy $S(E|A)=S(B|A)$, being the second part of the Holevo bound $\chi_{AE}$, can be obtained from the respective conditional matrix of mode $B$ conditioned on the measurement on mode $A$:
\begin{equation}
\gamma_{B|A}'=\left( \begin{array}{cc}
\eta(1/V_A+\epsilon)+1-\eta & 0 \\
0 & \eta(V_A+\epsilon)+1-\eta
\end{array} \right)
\end{equation}

However, for the entropy $S(E|A)$ in the case of a coherent-state protocol, the structure of the measurement at Alice must be taken into account, so the equality $S(E|A)=S(BC|A)$ holds, and the entropy must be calculated from the matrix of modes $BC$ conditioned on the $x$-measurement on mode $A$:
\begin{equation}
\gamma_{BC|A}'=\left( \begin{array}{cccc}
1+\eta\epsilon & 0 & -\frac{\sqrt{2\eta}(V_A-1)}{\sqrt{V_A^2-1}} & 0 \\
0 & \eta(V_A+\epsilon)+1-\eta & 0 & \frac{\sqrt{\eta(V_A^2-1)}}{\sqrt{2}} \\
-\frac{\sqrt{2\eta}(V_A-1)}{\sqrt{V_A^2-1}} & 0 & \frac{2V_A}{V_A+1} & 0 \\
0 & \frac{\sqrt{\eta(V_A^2-1)}}{\sqrt{2}} & 0 & \frac{1+V_A}{2}
\end{array} \right)
\end{equation}

From the derived matrices, the key rate can be calculated numerically.

\begin{figure}[H]
\centering
\begin{subfigure}[b]{0.3\textwidth}
\centering
\includegraphics[width=1.1\textwidth]{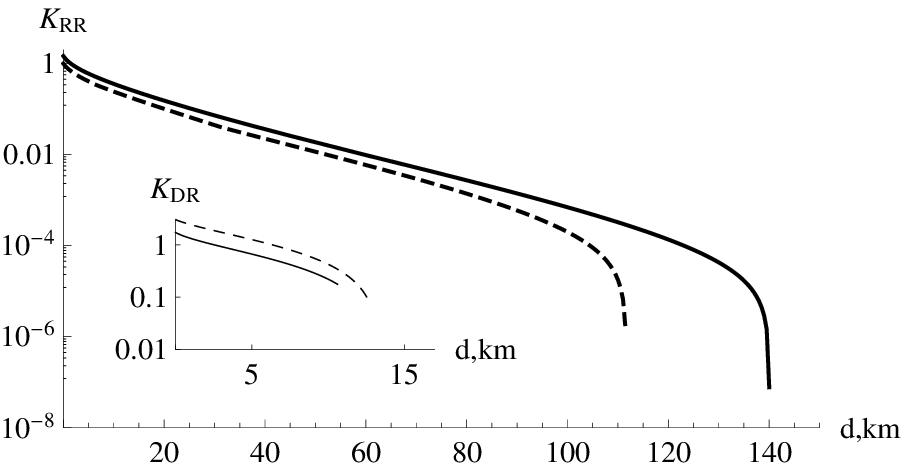}	
\caption{}
\end{subfigure}
\begin{subfigure}[b]{0.3\textwidth}
\centering
\includegraphics[width=1.1\textwidth]{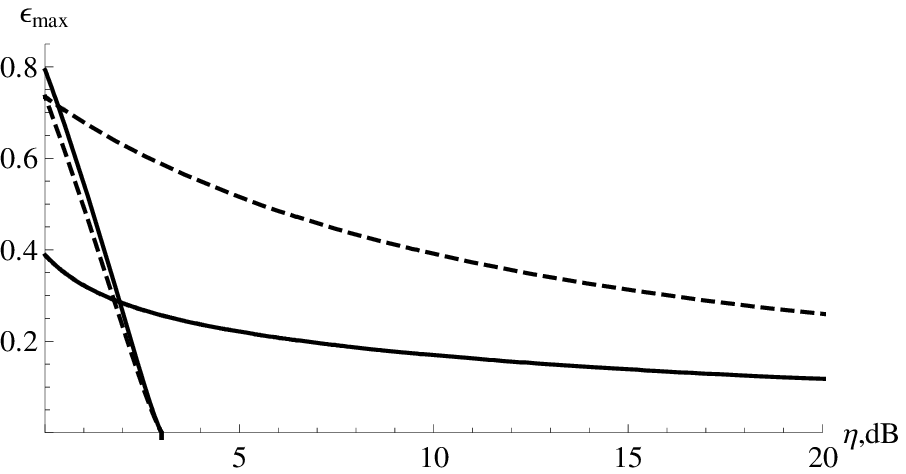}
\caption{}
\end{subfigure}
\begin{subfigure}[b]{0.3\textwidth}
\centering
\includegraphics[width=1.1\textwidth]{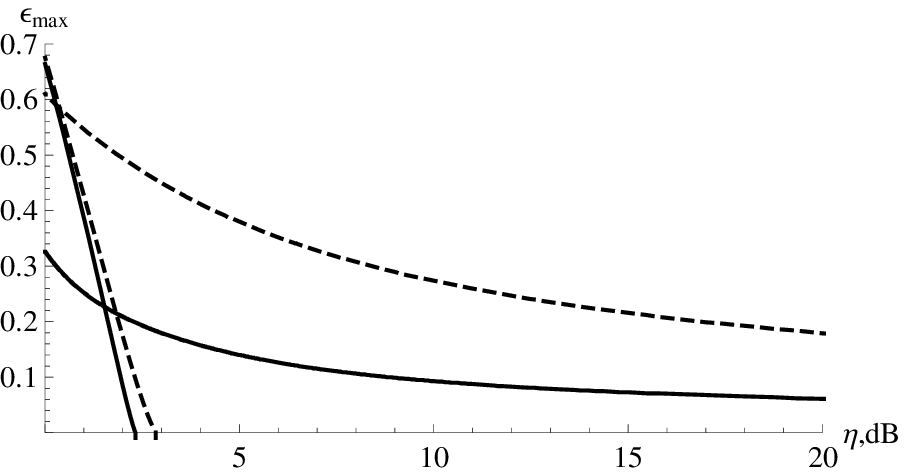}
\caption{}
\end{subfigure}
\caption{({\bf a}) Lower bound on secure key rate against channel distance in the telecom fiber with loss of $-$0.2 dB/km for {direct reconciliation} (DR) (main graph) and {reverse reconciliation} (RR) (inset) with imperfect post-processing, finite squeezing $V_S=$~0.1 and optimized modulation; ({\bf b}) maximum tolerable channel excess noise with respect to channel loss in $dB$ for the idealized case with perfect post-processing, infinite squeezing and arbitrarily strong modulation; ({\bf c})~maximum tolerable channel excess noise with respect to channel loss in $dB$ for the realistic case with imperfect post-processing, finite squeezing $V_S=0.1$ and optimized modulation. In all of the graphs, solid lines are for the coherent-state protocol, and dashed lines are for the squeezed-state one. The imperfect post-processing is taken as $\beta=0.95$ for RR and $\beta=0.99$ for DR. The curves on the maximum tolerable channel excess noise plots, which vanish at (or below) 3 dB, correspond to DR; the rest refer to RR.
\label{performance}}
\end{figure}

The security of the protocols in terms of the positivity of the lower bound on the secure key rate is thus limited by the channel transmittance and channel noise, which have a joint negative impact on security. Therefore, the protocols are mainly studied in terms of the tolerable loss (or, equivalently, the secure distance, assuming a standard telecom fiber with $-$0.2 dB attenuation per kilometer) and tolerable channel excess noise. We illustrate the typical performance of the protocols in the regime of optimal modulation for the given $\beta=0.95$ for RR and $\beta=0.99$ for DR (since the SNR is typically higher as the DR is applicable only for low loss, and the post-processing is less demanding in the DR regime) in Figure \ref{performance}. It is evident that RR CV QKD provides security at much stronger values of attenuation, while DR CV QKD can tolerate higher noise in the low-loss channels. The squeezed-state protocol shows better performance and robustness to noise in the RR case both for the perfect and imperfect post-processing. Interestingly, in the regime of non-ideal post-processing, the squeezed-state protocol also becomes superior to the coherent-state one in the DR case.

The above-described P \& M CV QKD protocol with coherent states was realized using homodyne detection by Lodewyck \emph{et al.} \cite{Lodewyck2007} and by Jouguet \emph{et al.} \cite{Jouguet2013} and using heterodyne detection (corresponding to the no-switching regime) by Lance \emph{et al.} \cite{Lance2005}.

Besides the fixed (fiber-type) channels, where transmittance is stable and excess noise is relatively low, the Gaussian CV QKD protocols were shown potentially applicable in the free-space atmospheric channels, where transmittance fluctuations due to atmospheric turbulence lead to additional excess noise, which can be however tolerated or compensated by post-selection \cite{Usenko2012}. 

The CV QKD protocols were also studied with respect to the imperfections of the trusted devices. First, the real homodyne detectors are inclined to non-unity detection efficiency and electronic noise, which altogether corrupt the data obtained by Bob from the homodyne measurement. The trusted {detection noise} in the coherent-state RR CV QKD protocol was first taken into account by Lodewyck \emph{et al.} \cite{Lodewyck2007} and later considered in the squeezed-state RR protocol by Garcia-Patron and Cerf \cite{Garcia2009}, who had shown that the trusted detection noise can even be helpful to provide robustness against the channel noise. It was also shown by Fossier~\emph{et~al.}~\cite{Fossier2009a} that the negative impact of imperfect detectors in the RR coherent-state CV QKD can be compensated using optical~pre-amplifiers.

On the trusted sender side, the state generation and modulation can also be imperfect, which results in the excess noise added to the signal at the stage of the state preparation. The preparation noise can therefore be the noise of the signal state itself (e.g., if a thermal state is produced by a noisy laser instead of a coherent one) or the noise added by an imperfect modulator. Such {preparation noise} was first addressed by Filip \cite{Filip2008} and shown harmful for the RR coherent-state CV QKD protocol for purely attenuating channels and then extended by Usenko and Filip to the noisy channels \cite{Usenko2010a}. It was also shown that the preparation noise can be purified by attenuating the signal (using, e.g., a simple variable beam splitter) prior to sending it to the channel. The method allows reduction (up to complete elimination in the regime of strong modulation) of the preparation noise. Later, it was shown by Weedbrook \emph{et al.} \cite{Weedbrook2010,Weedbrook2012} that the preparation noise can be tolerable and even helpful for the security of the DR coherent-state CV QKD protocol in the noisy channels. 

The role of other practical imperfections in coherent-state CV QKD was analyzed by Jouguet \emph{et al.} \cite{Jouguet2012}, who studied the imperfect realistic Gaussian modulation, the calibration of the detectors and the trusted phase noise in the sender station, and it was shown that these effects should be taken into account in the security analysis, especially in the finite-size regime. It was also shown that CV QKD systems can be in principle compromised using a wavelength attack on the local oscillator in the heterodyne \cite{Huang2013,Ma2013} and homodyne~\cite{Huang2014} detection, which can be however prevented by spectral filtering or real-time monitoring of shot noise \cite{Kunz2015}, the latter being also useful against the calibration attacks on the clock pulses in coherent-state CV QKD \cite{Jouguet2013a}. Furthermore, the fluctuations of the local oscillator were shown to be potentially harmful in CV QKD \cite{Ma2013a}, which can be compensated by tuning and monitoring of the intensity of the local oscillator by the trusted parties~\cite{Ma2014}. 

\section{Trusted Preparation and Detection Noise in CV QKD}

\begin{figure}[H]
\centering
\begin{subfigure}[b]{0.49\textwidth}
\centering
\includegraphics[width=1\textwidth]{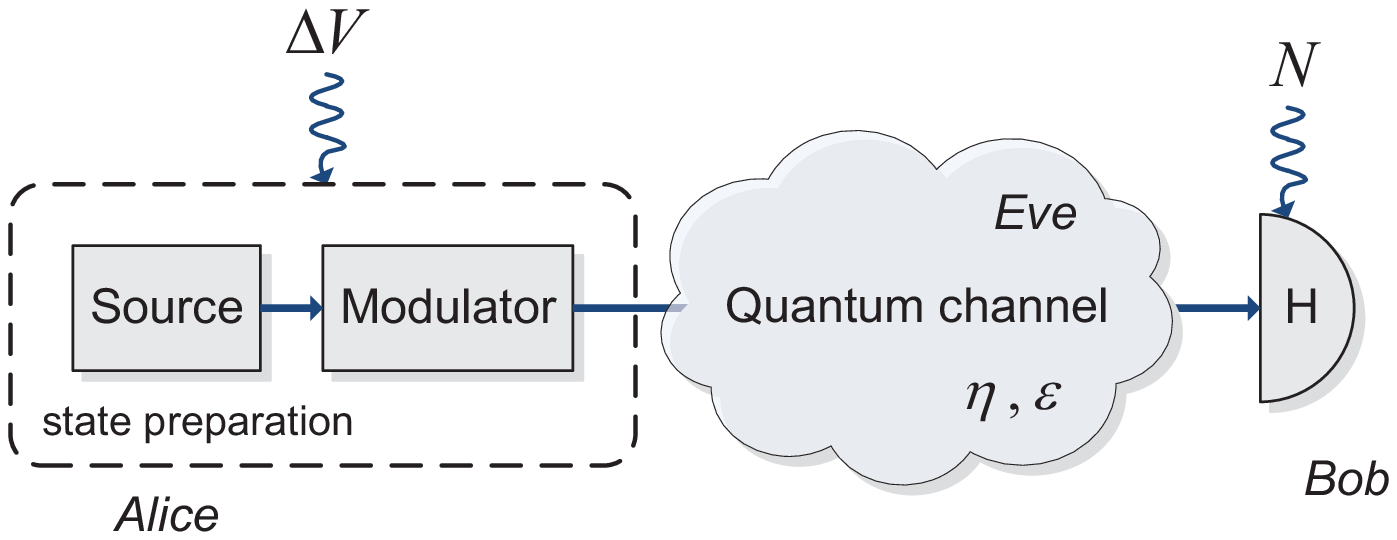}	
\caption{}
\end{subfigure}
\begin{subfigure}[b]{0.49\textwidth}
\centering
\includegraphics[width=1\textwidth]{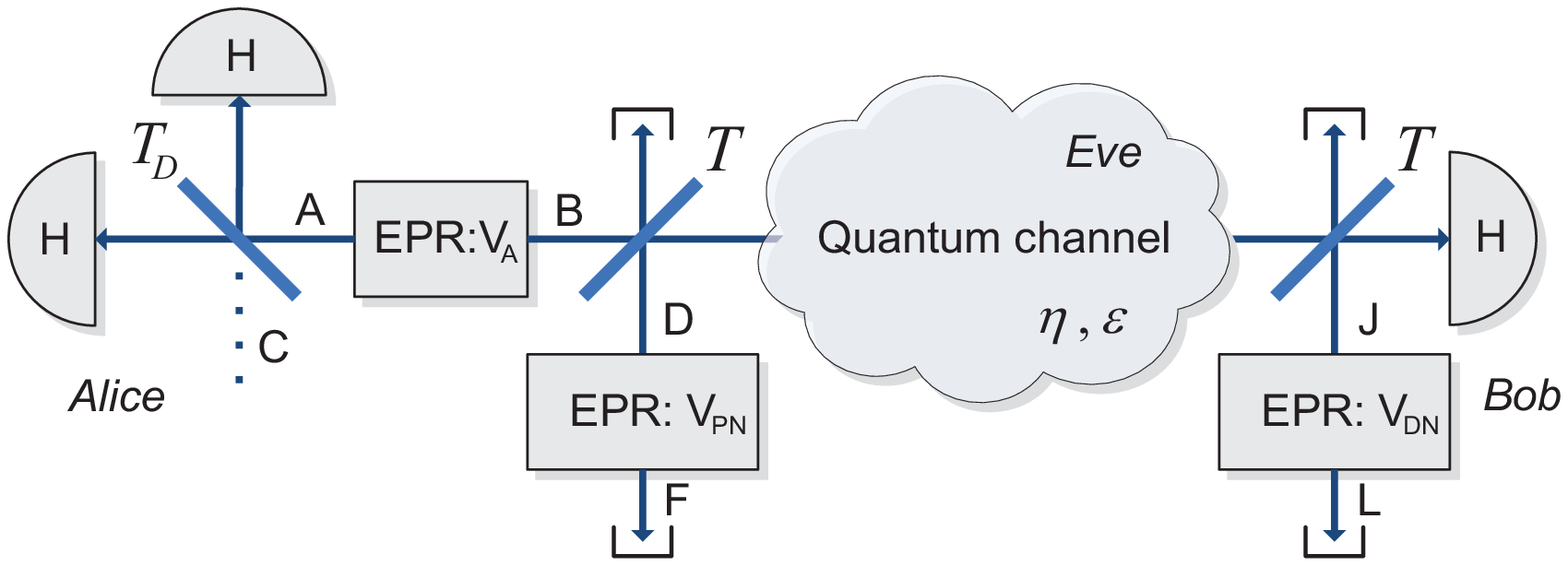}
\caption{}
\end{subfigure}
\caption{({\bf a}) CV QKD P \& M scheme with trusted preparation excess noise with variance $\Delta V$, affecting the state preparation, and trusted detection excess noise with variance $N$, which affects the homodyne detector; ({\bf b}) Equivalent EPR-based representation of the P \& M scheme used for purification of the trusted noise, which is done by introducing entangled sources with variances $V_{PN}$ and $V_{DN}$ on the sender and receiver side, respectively, and coupling one of the modes of each of the entangled sources to the signal mode on a strongly unbalanced beam splitter with transmittance $T \to 1$ for the signal modes, which, under proper setting of variances $V_{PN}$ and $V_{DN}$, becomes equivalent to the lossless addition of trusted excess noise.
\label{scheme_trustnoise}}
\end{figure}

In the current paper, we complete the analysis of the role of trusted preparation and detection noise in CV QKD with coherent and squeezed states, generalizing and extending the previously-known results. We consider the phase-insensitive excess noise added to the signal prior to the channel (preparation noise with variance $\Delta V$) and added to the signal after the channel (detection noise with variance $N$), as shown in Figure \ref{scheme_trustnoise}a.

The preparation and detection types of noise in the P \& M scheme do not affect the data, which Alice imposes by the modulation, as well as the correlation between Alice and Bob, but they change the covariance matrix of the state measured by Bob to $\gamma_B=Diag(\eta[V_S+V_M+\epsilon+\Delta V]+1-\eta+N,\eta[1/V_S+V_M+\epsilon+\Delta V]+1-\eta+N)$. Therefore, the mutual information between Alice and Bob reads:
\begin{equation}
I_{AB}=\frac{1}{2}\log_2{\Bigg[1+\frac{\eta V_M}{\eta(V_S+\epsilon+\Delta V)+1-\eta+N}\Bigg]}
\end{equation}

It is evident that both the preparation and detection types of noise reduce the mutual information. However, the effect on the security of the protocols can be different, as we show in the following sections, since the trusted noise also affects the Holevo bounds, which upper limit Eve's information.

In the simplest case of a purely lossy channel, the covariance matrix of the state available to Eve for collective measurement reads $\gamma_E=Diag\big([1-\eta](V_S+V_M+\Delta V)+\eta,[1-\eta](1/V_S+V_M+\Delta V)+\eta\big)$, the correlation matrix $\sigma_{AE}$ is not changed, while the correlation matrix $\sigma_{BE}$ becomes affected by the preparation noise and reads $\sigma_{BE}=Diag(\sqrt{\eta(1-\eta)}[1-V_S-V_M-\Delta V],\sqrt{\eta(1-\eta)}[1-1/V_S-V_M-\Delta V])$. The lower bound on the key rate can be then calculated analytically from these covariance matrices and will be analyzed in the particular cases in the following section.

In the more general case of a noisy untrusted channel, when a purification method must be applied and an equivalent EPR-based scheme is used instead of the P \& M one, the trusted preparation and detection noise can be purified by introducing additional EPR sources of modes $DF$ and $JL$, respectively, and coupling one of the modes of each of the sources to the signal, as shown in Figure \ref{scheme_trustnoise}b. A similar approach was previously used to purify the preparation \cite{Usenko2010a} and detection noise \cite{Lodewyck2007} in CV QKD purification-based security analysis. To losslessly add the trusted excess noise on the preparation and detection stage, the transmittance of the beam splitters used to couple the noisy modes with the signal should be made very high $T \to 1$. Then, after setting variances of the EPR-sources to $V_{PN}=\Delta V/(1-T)$ for the preparation noise and $V_{DN}=N/(1-T)$ for the detection noise, the coupling to EPR modes becomes numerically equivalent to the phase-insensitive trusted preparation excess noise $\Delta V$ and detection excess noise $N$, respectively. Alternatively, purification of the preparation noise (as well as of the detection noise) can be held by a third party \cite{Shen2011,Huang2013a}, which however gives the same result as the lossless coupling to an EPR mode. 

The calculation of the Holevo bounds in the case of noisy channels is based on the purity of the multimode state shared between the trusted parties and the fact that Eve holds the purification of the channel noise, so that $S(E)=S(ABDFJL)$, $S(E|B)=S(ADFJL|B)$ and $S(E|A)=S(BDFJL|A)$ for squeezed-state protocol and $S(E|A)=S(BCDFJL|A)$ for the coherent-state protocol. The overall covariance matrix of the state of the modes $ABDFJL$ after the propagation through the channel has the form:
\begin{equation}
\gamma_{ABDFJL}'=\left( \begin{array}{cccccc}
V_A\mathbb{I} & \tilde{C}_{AB}\sigma_z & C_{AD}\sigma_z & 0 & C_{AJ}\sigma_z & 0 \\
\tilde{C}_{AB}\sigma_z & \tilde{V}_B\mathbb{I} & C_{BD}\mathbb{I} & C_{BF}\sigma_z & C_{BJ}\mathbb{I} & C_{BL}\sigma_z \\
C_{AD}\sigma_z & C_{BD}\mathbb{I} & \tilde{V}_D & C_{DF}\sigma_z & C_{DJ}\mathbb{I} & 0 \\
0 & C_{BF}\sigma_z & C_{DF}\sigma_z & V_{PN}\mathbb{I} & C_{FJ}\sigma_z & 0 \\
C_{AJ}\sigma_z & C_{BJ}\mathbb{I} & C_{DJ}\mathbb{I} & C_{FJ}\sigma_z & \tilde{V}_J\mathbb{I} & C_{JL}\sigma_z \\
0 & C_{BL}\sigma_z & 0 & 0 & C_{JL}\sigma_z & V_{DN}\mathbb{I}
\end{array} \right)
\label{eprdash2}
\end{equation}
\noindent where:
\begin{eqnarray}
&\tilde{C}_{AB}=T\sqrt{\eta(V_A^2-1)}, \nonumber \\
&C_{AD}=-\sqrt{(1-T)(V_A^2-1)}, \nonumber \\
&C_{AJ}=-\sqrt{\eta T(1-T)(V_A^2-1)}, \nonumber \\
&\tilde{V}_B=T(\eta[TV_A+(1-T)V_{PN}+\epsilon]+1-\eta)+(1-T)V_{DN}, \nonumber \\
&C_{BD}=T\sqrt{\eta(1-T)}(V_{PN}-V_A), \nonumber \\
&C_{BF}=\sqrt{\eta T(1-T)(V_{PN}^2-1)}, \nonumber \\
&C_{BJ}=\sqrt{T(1-T)}(V_{DN}-\eta[TV_A+(1-T)V_{PN}+\epsilon]-1+\eta), \\
&C_{BL}=\sqrt{(1-T)(V_{DN}^2-1)}, \nonumber \\
&\tilde{V}_D=TV_{PN}+(1-T)V_A, \nonumber \\
&C_{DF}=\sqrt{T(V_{PN}^2-1)}, \nonumber \\
&C_{DJ}=(1-T)\sqrt{\eta T}(V_A-V_{PN}), \nonumber \\
&C_{FJ}=-(1-T)\sqrt{\eta(V_{PN}^2-1)}, \nonumber \\
&\tilde{V}_J=TV_{DN}+(1-T)\Big[\eta[TV_A+(1-T)V_{PN}+\epsilon]+1-\eta\Big], \nonumber \\
&C_{JL}=\sqrt{T(V_{DN}^2-1)} \nonumber
\end{eqnarray}

The conditional matrices used for calculation of the conditional von Neumann entropies can be obtained using Equation (\ref{cvmatcond}). From these matrices, the Holevo bound and, respectively, the lower bound on the key rate in the case of collective attacks in a noisy channel can be obtained numerically and will be used in the analysis in the 
following sections.

%

\section{Trusted Noise as a Threat}

\subsection{Preparation Noise and Reverse Reconciliation}

It was previously shown that {the trusted preparation noise can break the security of the coherent-state RR CV QKD protocol} already for the purely lossy channel in the case of individual attacks \cite{Filip2008}, which was later extended on the noisy channels \cite{Usenko2010a}. Here, we generalize the result for the arbitrary pure signal states and show that {the tolerance to the preparation noise depends on the signal state squeezing}. Indeed, following the calculations given in the previous section, we can derive the lower bound on the key rate for RR in the case of collective attacks on the noisy channel, which in the limit of strong modulation ($V_M \to \infty$) converges to the simple modification of the expression Equation (\ref{limkrrrvs}):
\begin{equation}
K_{RR}^{V_M\to\infty}=\frac{1}{2}\Bigg[\log_2{\frac{1}{1-\eta}}-\log_2{\big(\eta [V_S+\Delta V]+1-\eta\big)}\Bigg]
\end{equation}

From this follows that even in the case of a purely-attenuating channel and perfect implementation of the RR CV QKD protocol, the security against collective attacks is bounded by the condition:
\begin{equation}
\Delta V < \frac{2-\eta}{1-\eta}-V_S
\end{equation}
which converges to the previously-known bound $\Delta V < 1/(1-\eta)$ for the coherent-state RR protocol~\cite{Filip2008}. Thus, in the limit of strong attenuation $\eta\to 0$, which is the region of interest for the RR protocols aimed at implementation upon strong loss compared to the DR ones, the coherent-state protocol can tolerate up to 1 SNU of excess preparation noise $\Delta V$, while the squeezed-state protocol with infinite squeezing $V_S \to 0$ would tolerate preparation excess noise $\Delta V= 2$ SNU, which illustrates another potential advantage of the squeezed-state RR CV QKD. However, much stronger preparation noise can be tolerated either upon higher channel transmittance (in the limit $\eta\to 1$, arbitrarily-strong preparation noise is tolerable by RR CV QKD protocols) or by using the noise filtering method (originally called purification by the authors), when the signal is attenuated prior to being sent to the channel \cite{Filip2008,Usenko2010a}, which optimally can completely remove the negative impact of preparation noise in RR CV QKD with infinite modulation or at least strongly suppress such impact in the realistic case \cite{Usenko2010a}. Furthermore, the monitoring of the source noise \cite{Yang2012} or noiseless amplification \cite{Wang2014a} can be applied to improve the security of CV QKD with noisy coherent states. Alternatively, DR CV QKD protocols can be used, since, as it was shown and as we recapitulate further, they are in principle robust against preparation noise \cite{Weedbrook2010,Weedbrook2012}.

The given amount of preparation noise effectively limits the tolerable channel loss, which can be seen from the reversed security bound in terms of the channel loss $\eta > 1-1/\Delta V$ for the coherent-state protocol or $\eta > 1-1/(\Delta V-1)$ for infinitely-squeezed states. This imposes a limitation on the channel transmittance in the otherwise perfect implementation starting from $\Delta V=1$ SNU for the coherent-state protocol and for $\Delta V=2$~SNU for the squeezed-state one with arbitrarily-strong squeezing. 

\begin{figure}[H]
\centering
\begin{subfigure}[b]{0.49\textwidth}
\centering
\includegraphics[width=1\textwidth]{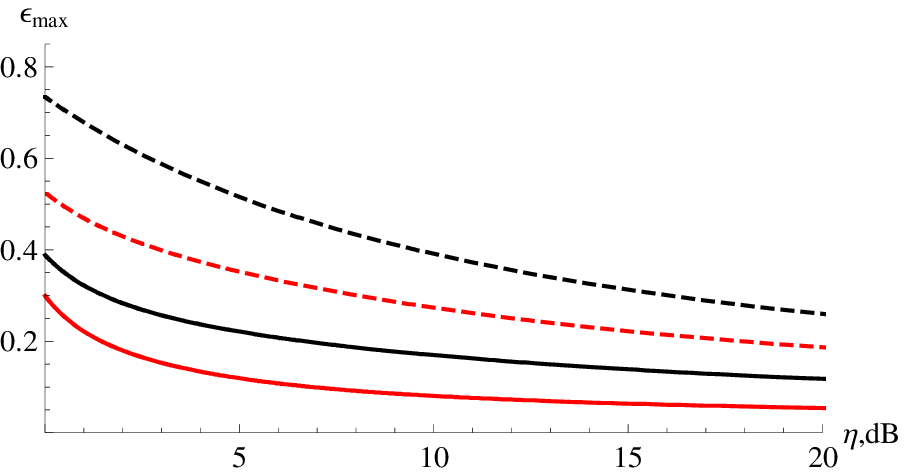}
\caption{}
\end{subfigure}	
\begin{subfigure}[b]{0.49\textwidth}
\centering
\includegraphics[width=1\textwidth]{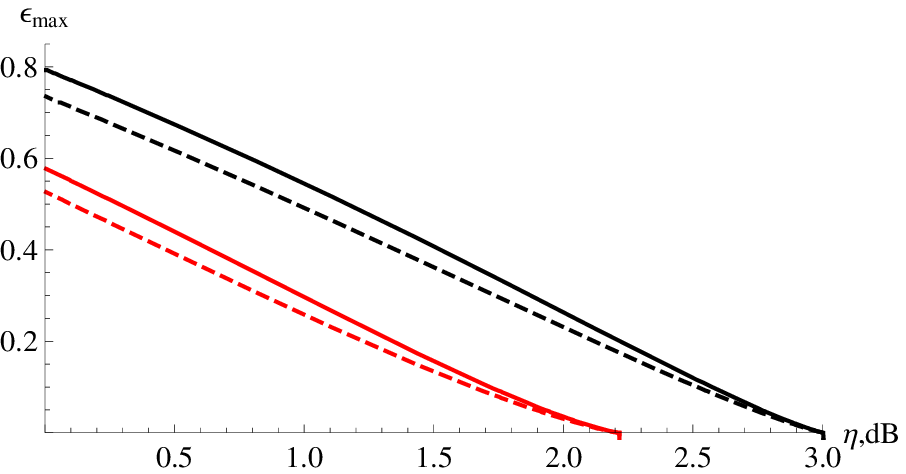}
\caption{}
\end{subfigure}
\caption{Maximum tolerable channel excess noise with respect to channel loss in $dB$ for the idealized case with perfect post-processing and arbitrarily strong modulation for the coherent-state protocol (solid lines) and the squeezed-state protocol with infinite squeezing (dashed lines). ({\bf a}) RR upon perfect implementation (black, upper lines) or in the presence of preparation noise $\Delta V=0.5$ shot-noise units (SNU) (red, lower lines);
({\bf b}) DR upon perfect implementation (black, upper lines) or in the presence of detection noise $N=0.5$ SNU (red, lower lines).
\label{epsmax_noise}}
\end{figure}

The preparation noise also reduces the robustness of the protocol to the channel noise, as can be seen in Figure \ref{epsmax_noise}a, where the tolerable channel excess noise is given in the presence of preparation noise $\Delta V=$~0.5~SNU in comparison to the perfect implementation of the protocols.

The mechanism of the negative effect of preparation noise in the RR CV QKD is two-fold. Firstly, it reduces the mutual information; secondly, it also increases the Holevo bound, \emph{i.e.}, the upper bound on the information leakage, since this type of noise is added prior to the channel and contributes to Eve's information. The preparation noise thus increases both the von Neumann entropies contributing to the Holevo bound, $S(E)$, as well as $S(E|B)$; however, the first one grows faster, and the Holevo bound is increased.

In the practical situations of limited modulation and other imperfections the RR CV QKD protocol is even more sensitive to the preparation noise; we will consider the effect of imperfections jointly in the Section \ref{combined}.

\subsection{Detection Noise and Direct Reconciliation}

While the {DR CV QKD protocols} are robust to the preparation noise, they {are sensitive to the detection noise, which can lead to a security break}, as we show here. Indeed, already in the limit of arbitrarily-strong modulation and perfect implementation of the protocol, when the key rate converges to the simple modification of Equation (\ref{limkrdrvs}) as:
\begin{equation}
K_{DR}^{V_M\to\infty}=\frac{1}{2}\Bigg[\log_2{\frac{\eta}{1-\eta}}-\log_2{\frac{V_S\eta+1-\eta+N}{V_S(1-\eta)+\eta}}\Bigg]
\end{equation}
the security is bounded by the condition:

\begin{equation}
N < \frac{2\eta-1}{1-\eta}
\end{equation}
which does not depend on the state squeezing $V_S$. The bound converges to zero, which means that any amount of detection noise cannot be tolerated by DR CV QKD, when the loss approaches 50\% (being the loss limit for DR protocols). On the other hand, reciprocally to the preparation noise in the RR case, large amounts of detection noise can be tolerated by the protocol, when the channel approaches lossless transmission ($\eta\to 1$). In the non-ideal case, however, detection noise can be harmful and must be seriously taken into account in any implementation of the DR protocols. Moreover, such noise quickly (faster than preparation noise in the RR case) limits the tolerable channel loss already for the otherwise perfect implementation of the protocols ($\beta=1$, $\epsilon=0$, $V_M\to\infty$). Indeed, the bound on the detection noise can be reversed to the bound on the channel loss being $\eta > 1-1/(N+2)$. That means that, e.g., for 1 SNU of detection noise, the transmittance even for the purely lossy channel should be no less than 66\%. 

In the presence of channel noise, similarly to the RR and preparation noise, the detection noise also reduces the robustness of the DR protocol to the channel noise, as can be seen in Figure \ref{epsmax_noise}b, where the tolerable channel excess noise is given in the presence of detection noise $N=0.5$ SNU in comparison to the perfect implementation of the protocols. Moreover, it can be seen that such an amount of detection noise already reduces the tolerable channel loss even for a purely attenuating channel.

Remarkably, the Holevo bound $\chi_{AE}$ is not sensitive to the detection noise; already in the purely lossy channel case, this noise is not present in the covariance matrices $\gamma_E$ and $\gamma_{E|A}$ and, thus, does not contribute to the information leakage (similarly for in the presence of the channel noise). Therefore, the security break due to detection noise in DR CV QKD is caused only by reduction of the mutual information $I_{AB}$ between the trusted~parties.

However, the trusted noise not only breaks the security, but can also improve the protocols, as we recapitulate and show in the next section.

\section{Trusted Noise as a Defense}

Despite the fact that trusted noise reduces the mutual information between the trusted parties, it can also, if applied at the reference side of the protocol, effectively decrease the information possessed by an eavesdropper on the data of the respective trusted party, that is decrease the Holevo quantity, upper bounding Eve's information. In the conditions of noisy channels, such a decrease can improve the key rate and extend the bounds of the tolerable channel noise, which is also known as {``fighting noise with noise''}. Further, we describe such effects for DR and RR protocols.

\subsection{Preparation Noise and Direct Reconciliation}

It was already shown that {the preparation noise} not only can be tolerated by the coherent-state DR CV QKD protocol \cite{Weedbrook2010,Weedbrook2012} (at least upon perfect implementation with $\beta=1$), but also {improves the robustness of the DR protocols to the channel noise}. Here, we generalize the result for the arbitrarily signal states and show the levels and the mechanisms of such an improvement.

\begin{figure}[H]
\centering
\begin{subfigure}[b]{0.32\textwidth}
\centering
\includegraphics[width=1\textwidth]{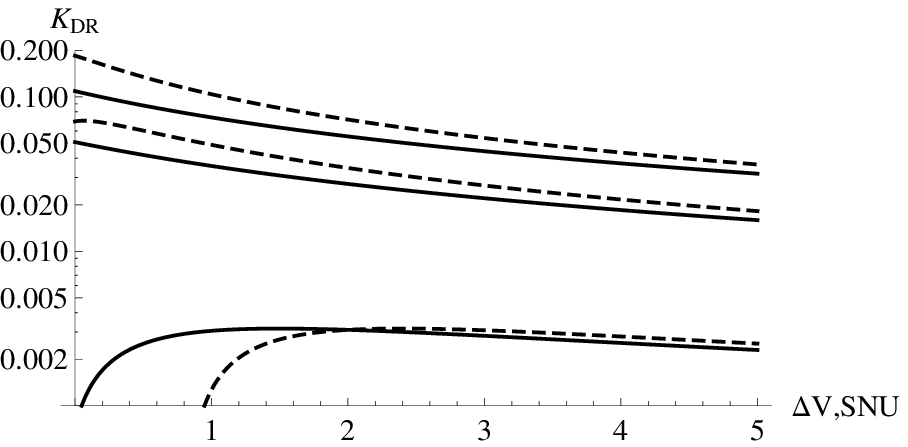}
\caption{}
\end{subfigure}	
\begin{subfigure}[b]{0.32\textwidth}
\centering
\includegraphics[width=1\textwidth]{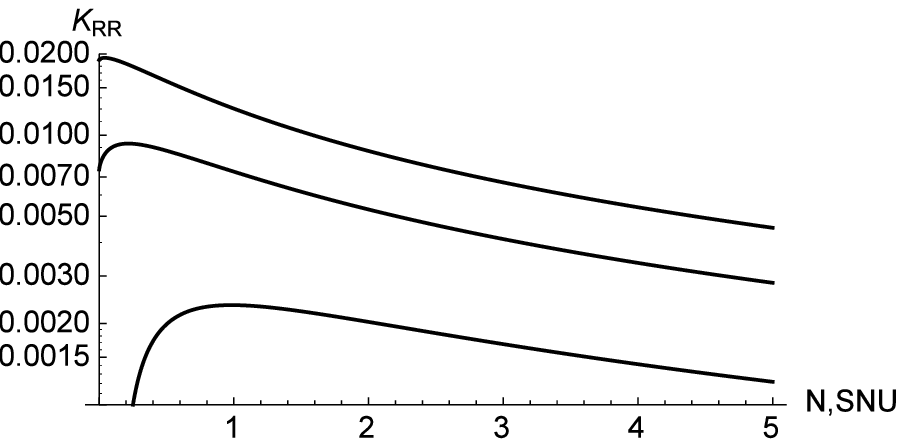}
\caption{}
\end{subfigure}	
\begin{subfigure}[b]{0.32\textwidth}
\centering	
\includegraphics[width=1\textwidth]{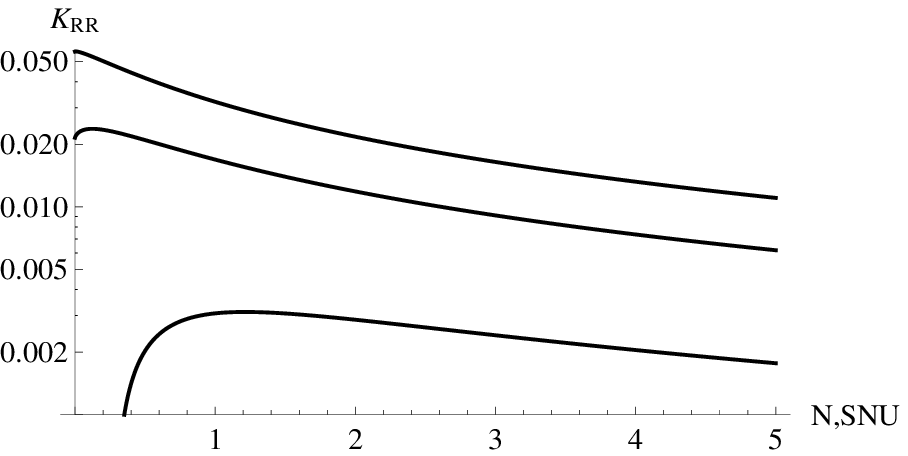}
\caption{}
\end{subfigure}	
\caption{Lower bound on the key rate in the case of collective attacks in a noisy channel on the protocols with arbitrarily-strong modulation $V_M\to\infty$ and perfect post-processing $\beta=1$. ({\bf a}) Key rate for the DR coherent-state (solid lines) and squeezed-state with $V_S=0.1$ (dashed lines) CV QKD protocols \textit{versus} preparation noise (in SNU) in the presence of channel noise $\epsilon=0.2,0.15,0.1$ (from bottom to top); the channel transmittance is $\eta=0.6$; ({\bf b}) key rate for RR coherent-state CV QKD protocol \textit{versus} detection noise (in SNU) in the presence of channel noise $\epsilon=0.18,0.15,0.12$ (from bottom to top); channel transmittance is $\eta=0.1$; ({\bf c}) key rate for the RR squeezed-state CV QKD protocol with $V_S=0.1$ \textit{versus} detection noise (in SNU) in the presence of channel noise $\epsilon=0.4,0.3,0.2$ (from bottom to top); channel transmittance is $\eta=0.1$. 
\label{KR_improvement}}
\end{figure}

We demonstrate the typical effect of the preparation noise on the lower bound on the key rate in the DR CV QKD in Figure \ref{KR_improvement}a for fixed channel transmittance and different amounts of the channel noise. When the channel noise is low, the preparation noise gradually reduces the key rate. As the channel noise increases, slight improvement becomes visible for low amounts of preparation noise. Close to the maximum tolerable channel noise; however, the preparation noise evidently improves the key rate and can even restore the security of the protocol. Similar behavior is observed for other channel transmittances. Interestingly, while the squeezed-state protocol is quantitatively superior to the coherent-state one already for feasible squeezing of $V_S=0.1$ at low channel noise, the situation becomes opposite close to the maximum tolerable channel noise. The presence of the preparation noise in this case makes coherent- and squeezed-state protocols quantitatively equivalent. Evidently, the preparation noise must be optimized to maximize the key rate as the channel noise gets stronger, while close to the maximum tolerable channel noise, the effect of preparation noise saturates. It is also evident that quantitative improvement is such a regime is very small, and one could expect that the improvement would be canceled by the finite-size effects, which reduce the key rate \cite{Leverrier2010,Ruppert2014}.

\begin{figure}[H]
\centering
\begin{subfigure}[b]{0.49\textwidth}
\centering
\includegraphics[width=1\textwidth]{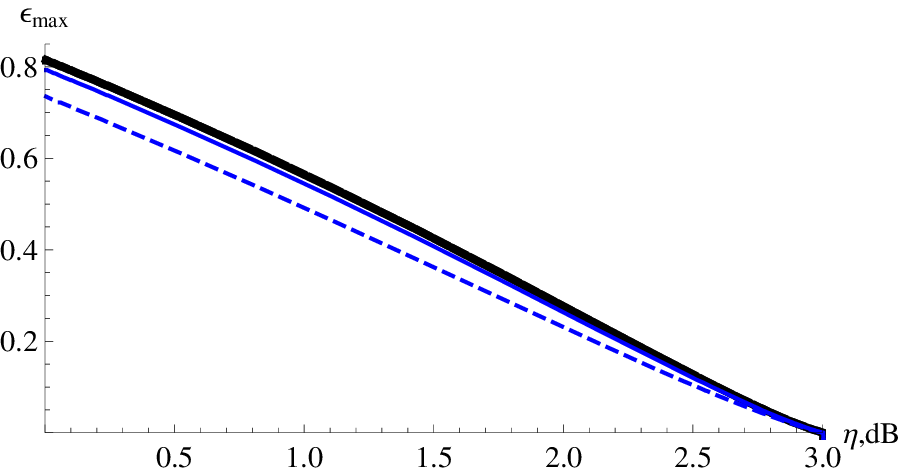}
\caption{}
\end{subfigure}	
\begin{subfigure}[b]{0.49\textwidth}
\centering
\includegraphics[width=1\textwidth]{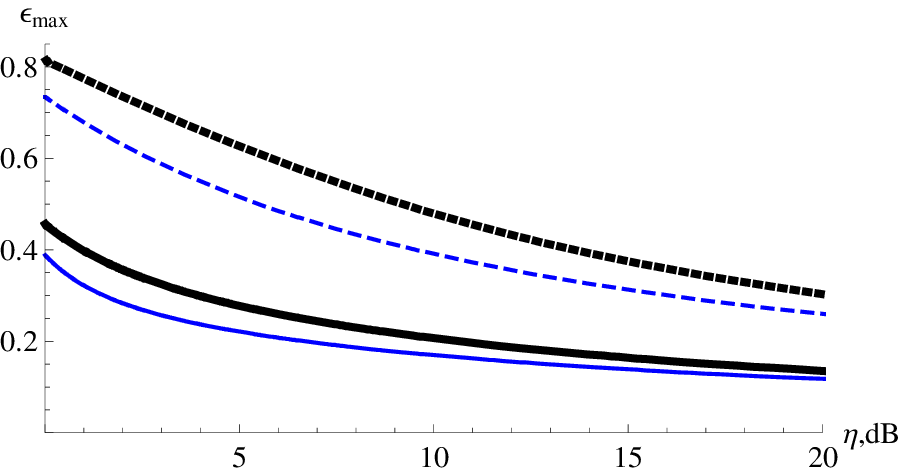}	
\caption{}
\end{subfigure}	
\caption{Maximum tolerable channel excess noise \textit{versus} channel transmittance (in dB scale) for the coherent-state (solid lines) and squeezed-state with $V_S=0.1$ (dashed lines) CV QKD protocols with arbitrarily-strong modulation $V_M\to\infty$ and perfect post-processing $\beta=1$. ({\bf a}) DR with no preparation noise (regular blue lines) and with optimal preparation noise added (thick black lines, overlapping for squeezed- and coherent-state protocols); ({\bf b}) RR with no preparation noise (regular blue lines) and with optimal detection noise added (thick black lines). 
\label{emax_improvement}}
\end{figure}

The improvement of the robustness to the channel noise by the preparation noise for DR CV QKD is evident from Figure \ref{emax_improvement}a. When the optimal amount of preparation noise is added, the difference between the coherent- and squeezed-state protocols vanishes. However, the improvement gets less as the channel loss~increases.

The reason for the improvement is concerned with the fact that the noise, added on the reference side of the protocol (the preparation noise in DR in this case), reduces the Holevo bound. Such noise, in fact, increases the both von Neumann entropies $S(E)$ and $S(E|A)$, but the latter one grows faster, thus contributing to the decrease of $\chi_{EA}$. Therefore, despite the simultaneous decrease of the mutual information $I_{AB}$, the lower bound on the key rate can be slightly improved and even turned positive by the preparation noise for strong channel noise. This also explains, why in the perfect conditions of noiseless implementation, the coherent-state protocol shows slightly better robustness to the channel noise compared to the squeezed-state protocol, which provides higher mutual information, but also offers more information leakage upon high channel noise. The shot noise of the coherent signal states in this case already reduces the Holevo bound compared to the squeezed signal state.

\subsection{Detection Noise and Reverse Reconciliation}

Similarly to the previous case, {the detection noise can improve the robustness of the RR CV QKD protocols to the channel noise}. This effect was already shown for the squeezed-state protocol \cite{Garcia2009}, while here, we generalize it for the arbitrary pure signal states and discuss the mechanism of such an improvement.

First, we show in Figure \ref{KR_improvement}b,c how the key rate can be improved by detection noise in the RR case for both the coherent- and squeezed-state protocols. Similarly to the case of DR and preparation noise, the detection noise decreases the key rate when the channel noise is relatively low, but can optimally improve the key rate and even restore the security of the protocols when
the channel noise is close to the threshold. Therefore, the detection noise must be optimized and can shift the bounds on the tolerable channel excess noise, as can be seen in Figure \ref{emax_improvement}b. In this case, however, the squeezed- and coherent-state protocols are both improved in their robustness to noise and do not overlap, \emph{i.e.}, the squeezed-state protocol remains to be more robust against noise, contrary to the DR case. This effect also explains why the RR protocols with the heterodyne detection at Bob's side are more stable against the channel noise, because an additional half SNU added at the detection stage in this case serves as the detection noise, improving the robustness of the protocol. However, typically, more noise must be added to achieve optimal performance in the noisy channels, at least in the asymptotic regime.

The reason for the improvement, similarly to DR and preparation noise, lays in the decrease of the Holevo bound, \emph{i.e.}, of the information leakage. While the von Neumann entropy $S(E)$ remains unchanged by the detection noise, the conditional entropy $S(E|B)$ increases (as Eve becomes effectively decoupled from Bob's data upon the addition of the uncorrelated noise) and $\chi_{BE}$ decreases. In particular, for the coherent-state protocol $V_S=1$, already in the purely-attenuating channel $\epsilon=0$ the derivative of $S(E|B)$ by $N$ at $N=0$ in the limit of large modulation $V_M \to \infty$ analytically simplifies to $(1-\eta)/(2\log{2})$, which is always non-negative. Such an increase of $S(E|B)$ improves the key rate when the channel noise is strong enough despite the decrease of the mutual information $I_{AB}$.

Note that the quantitative improvement of the key rate in the noisy channels is also minor as in the case of DR and preparation noise. Therefore, one could also expect that the positive effect of the detection noise would be canceled by the reduction of the key rate in the finite-size regime \cite{Leverrier2010,Ruppert2014}.

\section{Combined Trusted Noise Effects on Security}
\label{combined}

In the previous sections, we described the effects of trusted noise in the idealized DR and RR CV QKD protocols independently. However, the implementation of the protocols is never perfect, other different realistic effects may take place, and the types of trusted noise can be combined, which can amplify the negative and compensate the positive effects described above. Therefore, in the current section, we briefly study how the imperfect post-processing affects the performance of the protocols in the presence of trusted noise and what impact do combined sources of noise have on the security and robustness of the protocols.

\subsection{Trusted Noise and Imperfect Post-Processing}

First, {we {confirm the} positive effect of the trusted noise on the reference side of the protocols in the regime of limited post-processing efficiency}. Therefore, upon the realistic post-processing, the preparation noise in DR and the detection noise in RR (the latter effect previously shown in \cite{Madsen2012}) still improve the robustness of the protocols to the channel noise when the modulation variance $V_M$ is properly optimized for the given post-processing efficiency $\beta$ (values discussed above).

\begin{figure}[H]
\centering
\begin{subfigure}[b]{0.49\textwidth}
\centering
\includegraphics[width=1\textwidth]{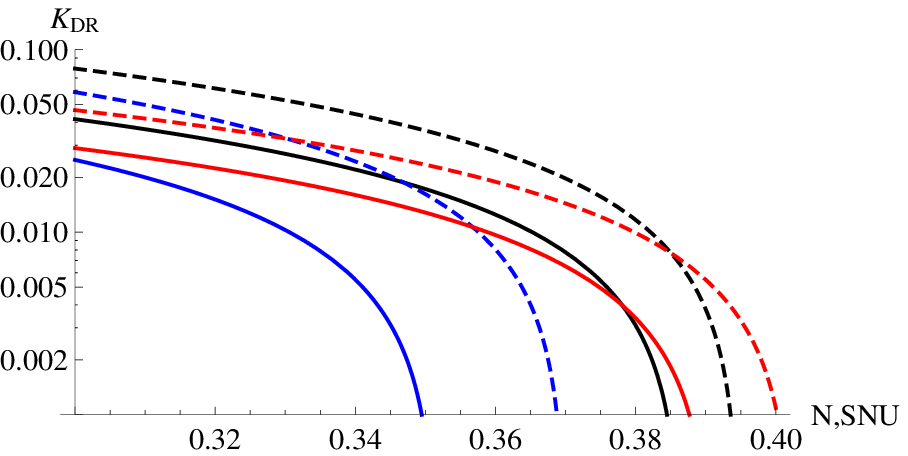}
\caption{}
\end{subfigure}	
\begin{subfigure}[b]{0.49\textwidth}
\centering
\includegraphics[width=1\textwidth]{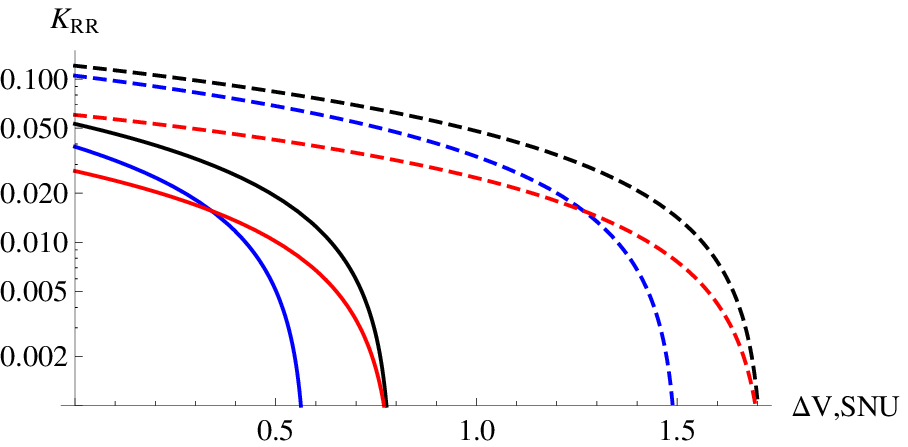}	
\caption{}
\end{subfigure}	
\caption{Lower bound on the key rate secure against collective attacks in a noisy channel on the protocols with limited modulation for coherent-state (solid lines) and squeezed-state with $V_S=0.1$ (dashed lines) CV QKD protocols. ({\bf a}) Key rate for DR \textit{versus} detection noise (in SNU) upon modulation variance $V_M=20$ in the perfect implementation ($\beta=1,$ $\Delta V=0$), black lines; upon limited post-processing efficiency ($\beta=$~0.99, $\Delta V$~=~0) blue lines; and in the presence of additional preparation noise $\Delta V=1$ SNU (red lines). Channel transmittance is $\eta=0.6$, channel noise is $1\%$ SNU. ({\bf b}) Key rate for RR \textit{versus} preparation noise (in SNU) upon modulation variance $V_M=5$ in the perfect implementation ($\beta=1$, $N=0$); black lines, upon limited post-processing efficiency ($\beta=0.95,$ $N=0$); blue lines; and in the presence of additional detection noise $N=1$ SNU (red lines). Channel transmittance is $\eta=0.1$; channel noise is $1\%$ SNU.
\label{KR_degrad}}
\end{figure}

On the other hand, the imperfect post-processing leads to a decrease of robustness to the trusted noise on the remote side of the protocols (that is to the detection noise in DR and preparation noise in RR). We illustrate the effect by the plots in Figure \ref{KR_degrad} (blue lines), where it is evident for both DR and RR protocols that the key rate becomes lower, and the security bound in terms of the noise on the remote side is reduced.

This effect is more significant for RR and is minor in the case of DR protocols, since the post-processing in the DR regime is typically higher, as well as the channel transmittance. 

\subsection{Combination of Preparation and Detection Noise}

Finally, we consider the combined effect of the trusted detection and preparation noise on the security of DR and RR CV QKD protocols. 

Interestingly, in this regime, the noise added on the reference side of the protocols can improve the robustness to the noise added on the remote side, \emph{i.e.}, the effects similar to ``fighting noise with noise'' take place, when one type of trusted noise improves the robustness against another. 

\begin{figure}[H]
\centering
\begin{subfigure}[b]{0.32\textwidth}
\centering
\includegraphics[width=1\textwidth]{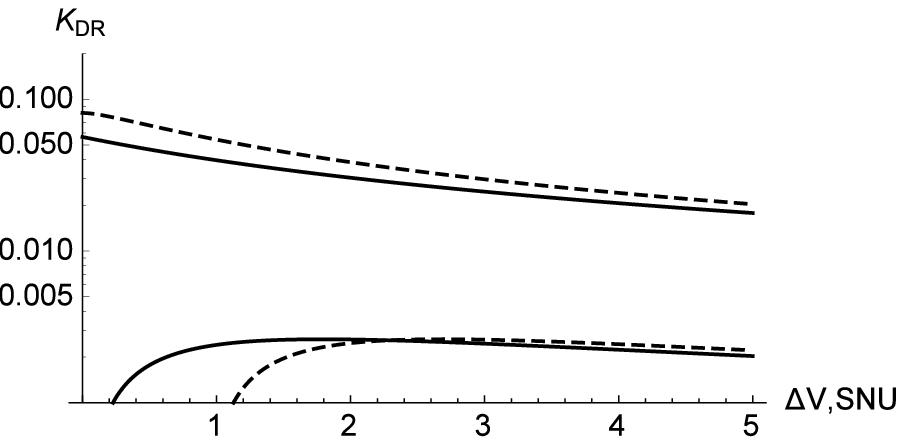}
\caption{}
\end{subfigure}	
\begin{subfigure}[b]{0.32\textwidth}
\centering
\includegraphics[width=1\textwidth]{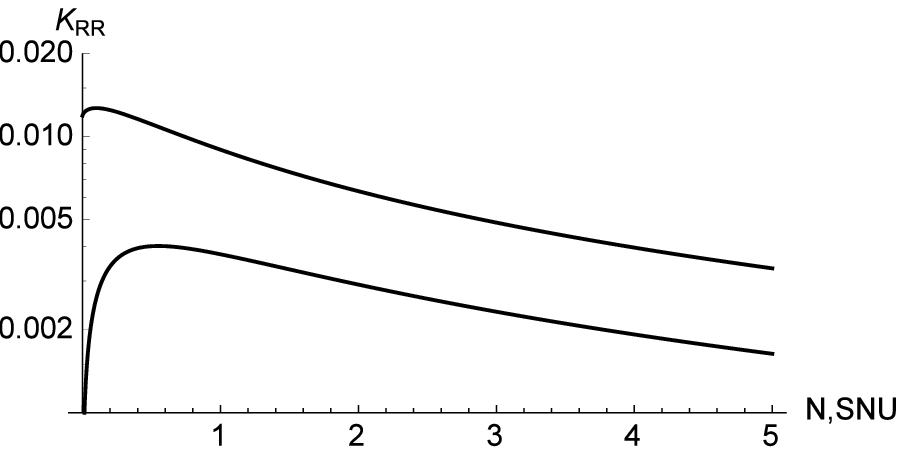}	
\caption{}
\end{subfigure}	
\begin{subfigure}[b]{0.32\textwidth}
\centering
\includegraphics[width=1\textwidth]{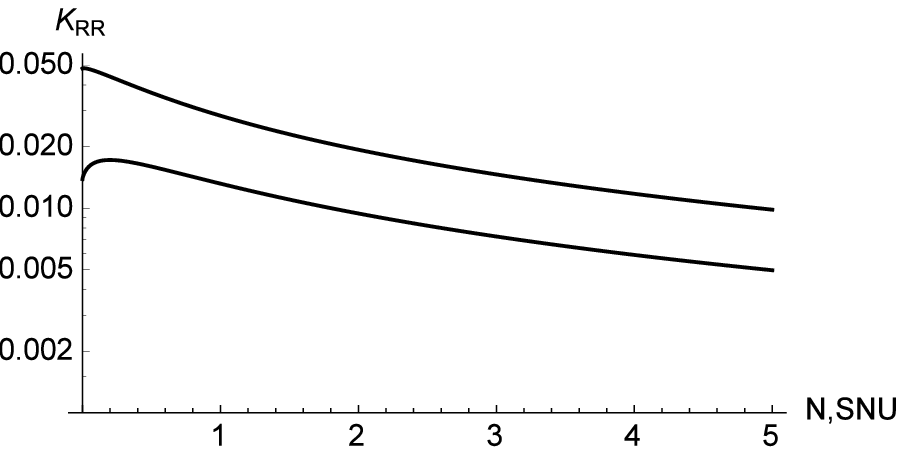}
\caption{}
\end{subfigure}	
\caption{Lower bound on the key rate in the case of collective attacks in a noisy channel on the protocols with arbitrarily-strong modulation $V_M\to\infty$ and perfect post-processing $\beta=1$. ({\bf a}) Key rate for DR coherent-state (solid lines) and squeezed-state with $V_S=0.1$ (dashed lines) CV QKD protocols \textit{versus} preparation noise (in SNU) in the presence of channel noise $\epsilon=0.15,$ $0.1$ (from bottom to top) and detection noise $N=8\%$ SNU, channel transmittance is $\eta=0.6$; ({\bf b}) key rate for RR coherent-state CV QKD protocol \textit{versus} detection noise (in SNU) in the presence of channel noise $\epsilon=0.12,$ $0.06$ (from bottom to top) and preparation noise $\Delta V=0.1$; channel transmittance is $\eta=0.1$; ({\bf c}) key rate for RR squeezed-state CV QKD protocol with $V_S=0.1$ \textit{versus} detection noise (in SNU) in the presence of channel noise $\epsilon=0.25,$ $0.1$ (from bottom to top) and preparation noise $\Delta V=0.1$; channel transmittance is $\eta=0.1$. 
\label{KR_double}}
\end{figure}

{In the case of DR protocols}, the addition of {trusted preparation noise slightly improves the robustness to the detection noise} (typically in the amount of fractions of a percent of SNU). The mechanism for such an improvement of robustness is similar to using noise on the reference side of the protocol against the channel noise: it leads to the decrease of the Holevo bound due to simultaneous increase of $S(E)$ and $S(E|A)$ entropies in the DR case with the latter one increasing faster. The effect gets less for the lower channel transmittance and upon the imperfect post-processing. This is shown in Figure \ref{KR_degrad}a (red lines): it is evident that the security region in terms of the detection noise is improved when additional preparation noise is present. We also confirm the positive effect of the preparation noise in the DR scheme in the presence of the detection noise, as shown in Figure \ref{KR_double}a, where the graphs for the key rate upon the same parameters as in Figure \ref{KR_improvement} are given in the presence of detection noise $N=8\%$ SNU (note that the lower graph corresponding to $\epsilon=0.2$ vanished in the presence of such detection noise). 

{In the case of RR protocols}, the addition of {trusted detection can result in improvement of the robustness to the trusted preparation noise}, which gets less for the lower channel transmittance. For the low transmittance, the effect is negligible, as we demonstrate in Figure \ref{KR_degrad}b,c (red lines), where the security bounds in terms of the preparation noise are not changed by the presence of the detection noise. At the same time, we confirm the positive role of the detection noise in the presence of the preparation noise, as can be seen from Figure~\ref{KR_double}b,c, where the key rate is plotted upon the same settings as in Figure \ref{KR_improvement}, but with additional preparation noise $\Delta V=0.1$. Note that, similarly to the DR case, the lines corresponding to the highest considered channel noise vanished in this case. Nevertheless, the positive effect of the detection noise is evident from the plots. Therefore, {the positive role of the trusted noise at the reference side of the protocols can be still observed in the presence of the trusted noise on the remote side}.

\section{Summary and Discussion}

We have recapitulated the known and shown the previously undisclosed effects of the phase-insensitive trusted preparation and detection noise in CV QKD. In particular, we have shown the overlapping saturation of the positive effect of the trusted preparation noise on the robustness of the DR protocols to the channel noise in the case of coherent and squeezed signal states. We have also shown the negative effect (up to security break in the purely attenuating channels) of the trusted detection noise in the DR scenario. We have confirmed the positive effect of the noise on the reference side of the protocols in the regime of imperfect post-processing. Finally we have shown that the combination of two types of trusted noise has practically no influence (and even slight improvement) on the security bounds on the harmful type of noise. We have discussed the mechanisms of the positive and negative influences of the trusted noise, which mainly concern the impact on the mutual information between the trusted parties, as well as on the quantum entropies, defining the information leakage for an imperfect quantum channel.

We have studied the P \& M schemes using the equivalent entanglement-based representation only to analyze the security in the case of the channel excess noise. The research, however, can be extended on the entangled-based protocols, which are promising for possible networking configurations and were recently studied in the alternative topology of entanglement in the middle by Weedbrook \cite{Weedbrook2013}. Such protocols were shown feasible on a proof-of-principle level by Su \emph{et al.} \cite{Su2009} without the additional modulation, and by Madsen~\emph{et~al.}~\cite{Madsen2012} with additional optimal modulation of the entangled states. They were also studied concerning the possible multi-mode effects \cite{Usenko2014,Usenko2015a}, which can affect the security of the CV QKD protocols with bright multimode (macroscopic) states of light \cite{Perina2007,Iskhakov2009}. 

Another promising direction of the study of the role of trusted noise in CV QKD is the two-way protocols suggested by Pirandola \emph{et al.} \cite{Pirandola2008a}, which are aimed at tolerating higher channel noise compared to the one-way CV QKD protocols discussed here. The trusted noise can have an impact on such protocols, and the role of trusted preparation noise on the two-way coherent-state CV QKD was recently discussed {by Weedbrook} \emph{et~al.}~\cite{Weedbrook2014} and by Wang \emph{et al.} \cite{Wang2014}, who had shown that the preparation noise tightens the bounds on the key rate, but preserves the advantage in terms of the tolerable channel noise of the two-way protocol over the one-way counterpart.

The physical discussion of the role of trusted preparation and detection noise in CV QKD is relevant not only for existing optical implementation of the protocols. The original motivation of the proposal of the DR CV QKD protocol with a noisy source \cite{Weedbrook2010} was to address the possibility of a short distance implementation of the protocol with noisy microwave sources. However, also the homodyne/heterodyne detection of the microwave states is typically noisy, and therefore, our analysis completes the description of the practical specific issues in the possible future implementations of CV QKD with microwave technology, which is rapidly \linebreak developing~\cite{Silva2010,Bozyigit2011,Eichler2011}.


\acknowledgments{\textbf{Acknowledgments:} The research leading to these results has received funding from the EU FP7 under Grant Agreement No. 308803 (Project BRISQ2), co-financed by {the Ministry of Education, Youth and Sports of the Czech Republic} (7E13032). Vladyslav C. Usenko acknowledges Project No. 13-27533J of~{the Czech Science Foundation} 
.}


\authorcontributions{\textbf{Author Contributions:} Vladyslav C. Usenko and Radim Filip  performed the theoretical calculations, discussed the results, prepared the manuscript, and commented on the manuscript at all stages. Both authors have read and approved the final manuscript.} 


\conflictofinterests{\textbf{Conflicts of Interest:} The authors declare no conflict of interest.} 

\bibliographystyle{mdpi}
\renewcommand\bibname{References}


\begin{thebibliography}{999}

\bibitem[Gisin \em{et~al.}(2002)Gisin, Ribordy, Tittel, and Zbinden]{Gisin2002}
Gisin, N.; Ribordy, G.; Tittel, W.; Zbinden, H.
\newblock Quantum cryptography.
\newblock {\em Rev. Mod. Phys.} {\bf 2002}, {\em 74}, 145--195.

\bibitem[Scarani \em{et~al.}(2009)Scarani, Bechmann-Pasquinucci, Cerf,
  Du{\v{s}}ek, L{\"u}tkenhaus, and Peev]{Scarani2009}
Scarani, V.; Bechmann-Pasquinucci, H.; Cerf, N.J.; Du{\v{s}}ek, M.;
  L{\"u}tkenhaus, N.; Peev, M.
\newblock The security of practical quantum key distribution.
\newblock {\em Rev. Mod. Phys.} {\bf 2009}, {\em 81},~1301--1350.

\bibitem[Vernam(1926)]{Vernam1926}
Vernam, G.S.
\newblock Cipher printing telegraph systems: For secret wire and radio
  telegraphic communications.
\newblock {\em J. AIEE } {\bf 1926}, {\em 45},~109--115.

\bibitem[Shannon(1949)]{Shannon1949}
Shannon, C.E.
\newblock Communication theory of secrecy systems.
\newblock {\em Bell Syst. Tech. J.} {\bf 1949}, {\em 28},~656--715.

\bibitem[Bennett and Brassard(1984)]{Bennett1984}
Bennett, C.H.; Brassard, G.
\newblock Quantum cryptography: Public Key Distribution and Coin Tossing.
\newblock  In~Proceedings of  the International Conference on Computers, Systems and Signal Processing, Bangalore, India, 10--19~December~1984.

\bibitem[Bennett(1992)]{Bennett1992}
Bennett, C.H.
\newblock Quantum cryptography using any two nonorthogonal states.
\newblock {\em Phys. Rev. Lett.} {\bf 1992}, {\em 68}, 3121--3124.

\bibitem[Scarani \em{et~al.}(2004)Scarani, Acin, Ribordy, and
  Gisin]{Scarani2004}
Scarani, V.; Acin, A.; Ribordy, G.; Gisin, N.
\newblock Quantum cryptography protocols robust against photon number splitting
  attacks for weak laser pulse implementations.
\newblock {\em Phys. Rev. Lett.} {\bf 2004}, {\em 92},~057901.

\bibitem[Brendel \em{et~al.}(1999)Brendel, Gisin, Tittel, and
  Zbinden]{Brendel1999}
Brendel, J.; Gisin, N.; Tittel, W.; Zbinden, H.
\newblock Pulsed energy-time entangled twin-photon source for quantum
  communication.
\newblock {\em Phys. Rev. Lett.} {\bf 1999}, {\em 82},~2594--2597.

\bibitem[Lo \em{et~al.}(2005)Lo, Ma, and Chen]{Lo2005}
Lo, H.K.; Ma, X.; Chen, K.
\newblock Decoy state quantum key distribution.
\newblock {\em Phys. Rev. Lett.} {\bf 2005}, {\em 94},~230504.

\bibitem[Einstein \em{et~al.}(1935)Einstein, Podolsky, and Rosen]{Einstein1935}
Einstein, A.; Podolsky, B.; Rosen, N.
\newblock Can quantum-mechanical description of physical reality be considered
  complete?
\newblock {\em Phys. Rev.} {\bf 1935}, {\em 47},~777--780.

\bibitem[Ekert(1991)]{Ekert1991}
Ekert, A.K.
\newblock Quantum cryptography based on Bell's theorem.
\newblock {\em Phys. Rev. Lett.} {\bf 1991}, {\em 67},~661--663.

\bibitem[Bell(1964)]{Bell1964}
Bell, J.S.
\newblock On the Einstein Podolsky Rosen paradox.
\newblock {\em Physics} {\bf 1964}, {\em 1},~195--200.

\bibitem[Clauser \em{et~al.}(1969)Clauser, Horne, Shimony, and
  Holt]{Clauser1969}
Clauser, J.F.; Horne, M.A.; Shimony, A.; Holt, R.A.
\newblock Proposed experiment to test local hidden-variable theories.
\newblock {\em Phys.~Rev. Lett.} {\bf 1969}, {\em 23},~880--884.

\bibitem[Bennett \em{et~al.}(1992)Bennett, Brassard, and Mermin]{Bennett1992a}
Bennett, C.H.; Brassard, G.; Mermin, N.D.
\newblock Quantum cryptography without Bell's theorem.
\newblock {\em Phys. Rev. Lett.} {\bf 1992}, {\em 68},~557--559.

\bibitem[Devetak and Winter(2005)]{Devetak2005}
Devetak, I.; Winter, A.
\newblock Distillation of secret key and entanglement from quantum states.
\newblock {\em  Proc. R. Soc. A} {\bf 2005}, {\em 461}, 207--235.

\bibitem[Kraus \em{et~al.}(2005)Kraus, Gisin, and Renner]{Kraus2005}
Kraus, B.; Gisin, N.; Renner, R.
\newblock Lower and upper bounds on the secret-key rate for quantum key
  distribution protocols using one-way classical communication.
\newblock {\em Phys. Rev. Lett.} {\bf 2005}, {\em 95},~080501.

\bibitem[M{\"u}ller-Quade and Renner(2009)]{Muller2009}
M{\"u}ller-Quade, J.; Renner, R.
\newblock Composability in quantum cryptography.
\newblock {\em New J. Phys.} {\bf 2009}, {\em 11},~085006.

\bibitem[Renner(2007)]{Renner2007}
Renner, R.
\newblock Symmetry of large physical systems implies independence of
  subsystems.
\newblock {\em Nat. Phys.} {\bf 2007}, {\em 3},~645--649.

\bibitem[Christandl \em{et~al.}(2009)Christandl, K{\"o}nig, and
  Renner]{Christandl2009}
Christandl, M.; K{\"o}nig, R.; Renner, R.
\newblock Postselection technique for quantum channels with applications to
  quantum cryptography.
\newblock {\em Phys. Rev. Lett.} {\bf 2009}, {\em 102},~020504.

\bibitem[Tomamichel \em{et~al.}(2012)Tomamichel, Lim, Gisin, and
  Renner]{Tomamichel2012}
Tomamichel, M.; Lim, C.C.W.; Gisin, N.; Renner, R.
\newblock Tight finite-key analysis for quantum cryptography.
\newblock {\em Nat.~Commun.} {\bf 2012}, {\em 3},~634, doi:10.1038/ncomms1631.

\bibitem[Braunstein and Van~Loock(2005)]{Braunstein2005}
Braunstein, S.L.; van~Loock, P.
\newblock Quantum information with continuous variables.
\newblock {\em Rev. Mod. Phys.} {\bf 2005}, {\em 77},~513--577, doi:10.1103/RevModPhys.77.513.

\bibitem[Weedbrook \em{et~al.}(2012)Weedbrook, Pirandola, Garcia-Patron, Cerf,
  Ralph, Shapiro, and Lloyd]{Weedbrook2012a}
Weedbrook, C.; Pirandola, S.; Garc\'ia-Patr\'on, R.; Cerf, N.J.; Ralph, T.C.;
  Shapiro, J.H.; Lloyd, S.
\newblock Gaussian quantum information.
\newblock {\em Rev. Mod. Phys.} {\bf 2012}, {\em 84}, doi:10.1103/RevModPhys.84.621.

\bibitem[Lorenz \em{et~al.}(2004)Lorenz, Korolkova, and Leuchs]{Lorenz2004}
Lorenz, S.; Korolkova, N.; Leuchs, G.
\newblock Continuous-variable quantum key distribution using polarization
  encoding and post selection.
\newblock {\em Appl. Phys. B} {\bf 2004}, {\em 79},~273--277.

\bibitem[Funk and Raymer(2002)]{Funk2002}
Funk, A.; Raymer, M.
\newblock Quantum key distribution using nonclassical photon-number
  correlations in macroscopic light pulses.
\newblock {\em Phys. Rev. A} {\bf 2002}, {\em 65},~042307.

\bibitem[Zhang \em{et~al.}(2003)Zhang, Kasai, and Hayasaka]{Zhang2003}
Zhang, Y.; Kasai, K.; Hayasaka, K.
\newblock Quantum channel using photon number correlated twin beams.
\newblock {\em Opt. Express} {\bf 2003}, {\em 11},~3592--3597.

\bibitem[Usenko and Lev(2005)]{Usenko2005}
Usenko, V.C.; Lev, B.I.
\newblock Large-alphabet quantum key distribution with two-mode coherently
  correlated beams.
\newblock {\em Phys.~Lett.~A} {\bf 2005}, {\em 348},~17--23.

\bibitem[Usenko and Paris(2007)]{Usenko2007}
Usenko, V.C.; Paris, M.G.
\newblock Multiphoton communication in lossy channels with photon-number
  entangled states.
\newblock {\em Phys.~Rev. A} {\bf 2007}, {\em 75},~043812.

\bibitem[Usenko and Paris(2010)]{Usenko2010}
Usenko, V.C.; Paris, M.G.
\newblock Quantum communication with photon-number entangled states and
  realistic photodetection.
\newblock {\em Phys. Lett. A} {\bf 2010}, {\em 374},~1342--1345.

\bibitem[Ralph(1999)]{Ralph1999}
Ralph, T.C.
\newblock Continuous variable quantum cryptography.
\newblock {\em Phys. Rev. A} {\bf 1999}, {\em 61},~010303.

\bibitem[Ralph(2000)]{Ralph2000}
Ralph, T.C.
\newblock Security of continuous-variable quantum cryptography.
\newblock {\em Phys. Rev. A} {\bf 2000}, {\em 62},~062306.

\bibitem[Hillery(2000)]{Hillery2000}
Hillery, M.
\newblock Quantum cryptography with squeezed states.
\newblock {\em Phys. Rev. A} {\bf 2000}, {\em 61},~022309.

\bibitem[Gottesman and Preskill(2001)]{Gottesman2001}
Gottesman, D.; Preskill, J.
\newblock Secure quantum key distribution using squeezed states.
\newblock {\em Phys. Rev. A} {\bf 2001}, {\em 63},~022309.

\bibitem[Reid(2000)]{Reid2000}
Reid, M.D.
\newblock Quantum cryptography with a predetermined key, using
  continuous-variable Einstein--Podolsky--Rosen correlations.
\newblock {\em Phys. Rev. A} {\bf 2000}, {\em 62},~062308.

\bibitem[Silberhorn \em{et~al.}(2002)Silberhorn, Korolkova, and
  Leuchs]{Silberhorn2002}
Silberhorn, C.; Korolkova, N.; Leuchs, G.
\newblock Quantum key distribution with bright entangled beams.
\newblock {\em Phys. Rev. Lett.} {\bf 2002}, {\em 88},~167902.

\bibitem[Cerf \em{et~al.}(2001)Cerf, Levy, and Van~Assche]{Cerf2001}
Cerf, N.J.; Levy, M.; van~Assche, G.
\newblock Quantum distribution of Gaussian keys using squeezed states.
\newblock {\em Phys. Rev. A} {\bf 2001}, {\em 63},~052311.

\bibitem[Grosshans and Grangier(2002)]{Grosshans2002}
Grosshans, F.; Grangier, P.
\newblock Continuous variable quantum cryptography using coherent states.
\newblock {\em Phys. Rev. Lett.} {\bf 2002}, {\em 88},~057902.

\bibitem[Silberhorn \em{et~al.}(2002)Silberhorn, Ralph, L{\"u}tkenhaus, and
  Leuchs]{Silberhorn2002a}
Silberhorn, C.; Ralph, T.C.; L{\"u}tkenhaus, N.; Leuchs, G.
\newblock Continuous variable quantum cryptography: Beating the 3~dB loss
  limit.
\newblock {\em Phys. Rev. Lett.} {\bf 2002}, {\em 89},~167901.

\bibitem[Grosshans \em{et~al.}(2003)Grosshans, Van~Assche, Wenger, Brouri,
  Cerf, and Grangier]{Grosshans2003}
Grosshans, F.; van~Assche, G.; Wenger, J.; Brouri, R.; Cerf, N.J.; Grangier, P.
\newblock Quantum key distribution using Gaussian-modulated coherent states.
\newblock {\em Nature} {\bf 2003}, {\em 421},~238--241.

\bibitem[Weedbrook \em{et~al.}(2004)Weedbrook, Lance, Bowen, Symul, Ralph, and
  Lam]{Weedbrook2004}
Weedbrook, C.; Lance, A.M.; Bowen, W.P.; Symul, T.; Ralph, T.C.; Lam, P.K.
\newblock Quantum cryptography without switching.
\newblock {\em Phys. Rev. Lett.} {\bf 2004}, {\em 93},~170504.

\bibitem[Usenko and Grosshans(2015)]{Usenko2015}
Usenko, V.C.; Grosshans, F.
\newblock Unidimensional continuous-variable quantum key distribution.
\newblock {\em Phys. Rev. A} {\bf 2015}, {\em 92},~062337.

\bibitem[Grosshans and Cerf(2004)]{Grosshans2004}
Grosshans, F.; Cerf, N.J.
\newblock Continuous-variable quantum cryptography is secure against
  non-Gaussian attacks.
\newblock {\em Phys.~Rev. Lett.} {\bf 2004}, {\em 92},~047905.

\bibitem[Navascu{\'e}s \em{et~al.}(2006)Navascu{\'e}s, Grosshans, and
  Acin]{Navascues2006}
Navascu{\'e}s, M.; Grosshans, F.; Acin, A.
\newblock Optimality of Gaussian attacks in continuous-variable quantum
  cryptography.
\newblock {\em Phys. Rev. Lett.} {\bf 2006}, {\em 97},~190502.

\bibitem[Garcia-Patron and Cerf(2006)]{Garcia2006}
Garc\'ia-Patr\'on, R.; Cerf, N.J.
\newblock Unconditional optimality of Gaussian attacks against
  continuous-variable quantum key distribution.
\newblock {\em Phys. Rev. Lett.} {\bf 2006}, {\em 97},~190503.

\bibitem[Wolf \em{et~al.}(2006)Wolf, Giedke, and Cirac]{Wolf2006}
Wolf, M.M.; Giedke, G.; Cirac, J.I.
\newblock Extremality of Gaussian quantum states.
\newblock {\em Phys. Rev. Lett.} {\bf 2006}, {\em 96},~080502.

\bibitem[Pirandola \em{et~al.}(2008)Pirandola, Braunstein, and
  Lloyd]{Pirandola2008}
Pirandola, S.; Braunstein, S.L.; Lloyd, S.
\newblock Characterization of collective Gaussian attacks and security of
  coherent-state quantum cryptography.
\newblock {\em Phys. Rev. Lett.} {\bf 2008}, {\em 101},~200504.

\bibitem[Pirandola \em{et~al.}(2009)Pirandola, Garc{\'\i}a-Patr{\'o}n,
  Braunstein, and Lloyd]{Pirandola2009}
Pirandola, S.; Garc{\'\i}a-Patr{\'o}n, R.; Braunstein, S.L.; Lloyd, S.
\newblock Direct and reverse secret-key capacities of a quantum channel.
\newblock {\em Phys. Rev. Lett.} {\bf 2009}, {\em 102},~050503.

\bibitem[Renner and Cirac(2009)]{Renner2009}
Renner, R.; Cirac, J.I.
\newblock de Finetti representation theorem for infinite-dimensional quantum
  systems and applications to quantum cryptography.
\newblock {\em Phys. Rev. Lett.} {\bf 2009}, {\em 102},~110504.

\bibitem[Braunstein and Kimble(1998)]{Braunstein1998}
Braunstein, S.L.; Kimble, H.J.
\newblock Teleportation of continuous quantum variables.
\newblock {\em Phys. Rev. Lett.} {\bf 1998}, {\em 80},~869--872, doi:10.1103/PhysRevLett.80.869.

\bibitem[Chizhov \em{et~al.}(2002)Chizhov, Kn{\"o}ll, and Welsch]{Chizhov2002}
Chizhov, A.; Kn{\"o}ll, L.; Welsch, D.G.
\newblock Continuous-variable quantum teleportation through lossy channels.
\newblock {\em Phys.~Rev.~A} {\bf 2002}, {\em 65},~022310.

\bibitem[Fiur{\'a}{\v{s}}ek(2002)]{Fiurasek2002}
Fiur{\'a}{\v{s}}ek, J.
\newblock Improving the fidelity of continuous-variable teleportation via local
  operations.
\newblock {\em Phys. Rev. A} {\bf 2002}, {\em 66},~012304.

\bibitem[Pirandola and Mancini(2006)]{Pirandola2006}
Pirandola, S.; Mancini, S.
\newblock Quantum teleportation with continuous variables: A survey.
\newblock {\em Laser Phys.} {\bf 2006}, {\em 16}, 1418--1438.

\bibitem[Cerf and Iblisdir(2000)]{Cerf2000}
Cerf, N.J.; Iblisdir, S.
\newblock Optimal {\it N}-to-{\it M} cloning of conjugate quantum variables.
\newblock {\em Phys. Rev. A} {\bf 2000}, {\em 62},~040301.

\bibitem[Braunstein \em{et~al.}(2001)Braunstein, Cerf, Iblisdir, van Loock, and
  Massar]{Braunstein2001}
Braunstein, S.L.; Cerf, N.J.; Iblisdir, S.; van Loock, P.; Massar, S.
\newblock Optimal cloning of coherent states with a linear amplifier and beam
  splitters.
\newblock {\em Phys. Rev. Lett.} {\bf 2001}, {\em 86},~4938--4941.

\bibitem[Fiur{\'a}{\v{s}}ek(2001)]{Fiurasek2001}
Fiur{\'a}{\v{s}}ek, J.
\newblock Optical implementation of continuous-variable quantum cloning
  machines.
\newblock {\em Phys. Rev. Lett.} {\bf 2001}, {\em 86}, 4942--4945.

\bibitem[Leverrier \em{et~al.}(2010)Leverrier, Grosshans, and
  Grangier]{Leverrier2010}
Leverrier, A.; Grosshans, F.; Grangier, P.
\newblock Finite-size analysis of a continuous-variable quantum key
  distribution.
\newblock {\em Phys.~Rev. A} {\bf 2010}, {\em 81},~062343.

\bibitem[Ruppert \em{et~al.}(2014)Ruppert, Usenko, and Filip]{Ruppert2014}
Ruppert, L.; Usenko, V.C.; Filip, R.
\newblock Long-distance continuous-variable quantum key distribution with
  efficient channel estimation.
\newblock {\em Phys. Rev. A} {\bf 2014}, {\em 90},~062310.

\bibitem[Leverrier and Grangier(2010)]{Leverrier2010a}
Leverrier, A.; Grangier, P.
\newblock Simple proof that Gaussian attacks are optimal among collective
  attacks against continuous-variable quantum key distribution with a Gaussian
  modulation.
\newblock {\em Phys. Rev. A} {\bf 2010}, {\em 81},~062314.

\bibitem[Walk \em{et~al.}(2013)Walk, Ralph, Symul, and Lam]{Walk2013}
Walk, N.; Ralph, T.C.; Symul, T.; Lam, P.K.
\newblock Security of continuous-variable quantum cryptography with Gaussian
  postselection.
\newblock {\em Phys. Rev. A} {\bf 2013}, {\em 87},~020303.

\bibitem[Fiur{\'a}{\v{s}}ek and Cerf(2012)]{Fiurasek2012}
Fiur{\'a}{\v{s}}ek, J.; Cerf, N.J.
\newblock Gaussian postselection and virtual noiseless amplification in
  continuous-variable quantum key distribution.
\newblock {\em Phys. Rev. A} {\bf 2012}, {\em 86},~060302.

\bibitem[Furrer \em{et~al.}(2012)Furrer, Franz, Berta, Leverrier, Scholz,
  Tomamichel, and Werner]{Furrer2012}
Furrer, F.; Franz, T.; Berta, M.; Leverrier, A.; Scholz, V.B.; Tomamichel, M.;
  Werner, R.F.
\newblock Continuous variable quantum key distribution: Finite-key analysis of
  composable security against coherent attacks.
\newblock {\em Phys. Rev. Lett.} {\bf 2012}, {\em 109},~100502.

\bibitem[Furrer(2014)]{Furrer2014}
Furrer, F.
\newblock Reverse-reconciliation continuous-variable quantum key distribution
  based on the uncertainty principle.
\newblock {\em Phys. Rev. A} {\bf 2014}, {\em 90},~042325.

\bibitem[Leverrier \em{et~al.}(2013)Leverrier, Garc{\'\i}a-Patr{\'o}n, Renner,
  and Cerf]{Leverrier2013}
Leverrier, A.; Garc{\'\i}a-Patr{\'o}n, R.; Renner, R.; Cerf, N.J.
\newblock Security of continuous-variable quantum key distribution against
  general attacks.
\newblock {\em Phys. Rev. Lett.} {\bf 2013}, {\em 110},~030502.

\bibitem[Leverrier(2015)]{Leverrier2015}
Leverrier, A.
\newblock Composable security proof for continuous-variable quantum key
  distribution with coherent states.
\newblock {\em Phys.~Rev.~Lett.} {\bf 2015}, {\em 114},~070501.

\bibitem[Heid and L{\"u}tkenhaus(2006)]{Heid2006}
Heid, M.; L{\"u}tkenhaus, N.
\newblock Efficiency of coherent-state quantum cryptography in the presence of
  loss: Influence of realistic error correction.
\newblock {\em Phys. Rev. A} {\bf 2006}, {\em 73},~052316.

\bibitem[Lodewyck \em{et~al.}(2007)Lodewyck, Bloch, Garc{\'\i}a-Patr{\'o}n,
  Fossier, Karpov, Diamanti, Debuisschert, Cerf, Tualle-Brouri, McLaughlin,
  et~al.]{Lodewyck2007}
Lodewyck, J.; Bloch, M.; Garc{\'\i}a-Patr{\'o}n, R.; Fossier, S.; Karpov, E.;
  Diamanti, E.; Debuisschert, T.; Cerf,~N.J.; Tualle-Brouri, R.; McLaughlin,
  S.W.; \emph{et al}.
\newblock Quantum key distribution over 25 km with an all-fiber
  continuous-variable system.
\newblock {\em Phys. Rev. A} {\bf 2007}, {\em 76},~042305.

\bibitem[Fossier \em{et~al.}(2009)Fossier, Diamanti, Debuisschert, Villing,
  Tualle-Brouri, and Grangier]{Fossier2009}
Fossier, S.; Diamanti, E.; Debuisschert, T.; Villing, A.; Tualle-Brouri, R.;
  Grangier, P.
\newblock Field test of a continuous-variable quantum key distribution
  prototype.
\newblock {\em New J. Phys.} {\bf 2009}, {\em 11},~045023.

\bibitem[Leverrier \em{et~al.}(2008)Leverrier, All\'eaume, Boutros, Z\'emor,
  and Grangier]{Leverrier2008}
Leverrier, A.; All\'eaume, R.; Boutros, J.; Z\'emor, G.; Grangier, P.
\newblock Multidimensional reconciliation for a~continuous-variable quantum key
  distribution.
\newblock {\em Phys. Rev. A} {\bf 2008}, {\em 77},~042325.

\bibitem[Jouguet \em{et~al.}(2011)Jouguet, Kunz-Jacques, and
  Leverrier]{Jouguet2011}
Jouguet, P.; Kunz-Jacques, S.; Leverrier, A.
\newblock Long-distance continuous-variable quantum key distribution with a~Gaussian modulation.
\newblock {\em Phys. Rev. A} {\bf 2011}, {\em 84},~062317.

\bibitem[Jouguet and Kunz-Jacques(2014)]{Jouguet2014}
Jouguet, P.; Kunz-Jacques, S.
\newblock High performance error correction for quantum key distribution using
  polar codes.
\newblock {\em Quantum Inf.  Comput.} {\bf 2014}, {\em
  14},~329--338.

\bibitem[Jouguet \em{et~al.}(2013)Jouguet, Kunz-Jacques, Leverrier, Grangier,
  and Diamanti]{Jouguet2013}
Jouguet, P.; Kunz-Jacques, S.; Leverrier, A.; Grangier, P.; Diamanti, E.
\newblock Experimental demonstration of long-distance continuous-variable
  quantum key distribution.
\newblock {\em Nat. Photonics} {\bf 2013}, {\em 7},~378--381.

\bibitem[Usenko and Filip(2011)]{Usenko2011}
Usenko, V.C.; Filip, R.
\newblock Squeezed-state quantum key distribution upon imperfect
  reconciliation.
\newblock {\em New J. Phys.} {\bf 2011}, {\em 13},~113007.

\bibitem[Grosshans(2005)]{Grosshans2005}
Grosshans, F.
\newblock Collective attacks and unconditional security in continuous variable
  quantum key distribution.
\newblock {\em Phys.~Rev.~Lett.} {\bf 2005}, {\em 94},~020504.

\bibitem[Grosshans \em{et~al.}(2003)Grosshans, Cerf, Wenger, Tualle-Brouri, and
  Grangier]{Grosshans2003a}
Grosshans, F.; Cerf, N.J.; Wenger, J.; Tualle-Brouri, R.; Grangier, P.
\newblock Virtual entanglement and reconciliation protocols for quantum
  cryptography with continuous variables.
\newblock {\em Quantum Inf. Comput.} {\bf 2003}, {\em 3},~535--552.

\bibitem[Navascu{\'e}s and Ac{\'\i}n(2005)]{Navascues2005}
Navascu{\'e}s, M.; Ac{\'\i}n, A.
\newblock Security bounds for continuous variables quantum key distribution.
\newblock {\em Phys. Rev. Lett.} {\bf 2005}, {\em 94},~020505.

\bibitem[Heid and L{\"u}tkenhaus(2007)]{Heid2007}
Heid, M.; L{\"u}tkenhaus, N.
\newblock Security of coherent-state quantum cryptography in the presence of
  Gaussian noise.
\newblock {\em Phys.~Rev.~A} {\bf 2007}, {\em 76},~022313.

\bibitem[Blandino \em{et~al.}(2012)Blandino, Leverrier, Barbieri, Etesse,
  Grangier, and Tualle-Brouri]{Blandino2012}
Blandino, R.; Leverrier, A.; Barbieri, M.; Etesse, J.; Grangier, P.;
  Tualle-Brouri, R.
\newblock Improving the maximum transmission distance of continuous-variable
  quantum key distribution using a noiseless amplifier.
\newblock {\em Phys. Rev. A} {\bf 2012}, {\em 86},~012327.

\bibitem[Mayers and Yao(1998)]{Mayers1998}
Mayers, D.; Yao, A.
\newblock Quantum Cryptography with Imperfect Apparatus.
\newblock  In Proceedings of the 39th
  Annual Symposium on Foundations of Computer Science (FOCS `98), Palo Alto, CA, USA,    8--11 November 1998.

\bibitem[Barrett \em{et~al.}(2005)Barrett, Hardy, and Kent]{Barrett2005}
Barrett, J.; Hardy, L.; Kent, A.
\newblock No signaling and quantum key distribution.
\newblock {\em Phys. Rev. Lett.} {\bf 2005}, {\em 95},~010503.

\bibitem[Ac{\'i}n \em{et~al.}(2006)Ac{\'i}n, Gisin, and Masanes]{Acin2006}
Ac{\'i}n, A.; Gisin, N.; Masanes, L.
\newblock From Bell's Theorem to Secure Quantum Key Distribution.
\newblock {\em Phys. Rev. Lett.} {\bf 2006}, {\em 97},~120405.

\bibitem[Ac{\'\i}n \em{et~al.}(2007)Ac{\'\i}n, Brunner, Gisin, Massar, Pironio,
  and Scarani]{Acin2007}
Ac{\'\i}n, A.; Brunner, N.; Gisin, N.; Massar, S.; Pironio, S.; Scarani, V.
\newblock Device-independent security of quantum cryptography against
  collective attacks.
\newblock {\em Phys. Rev. Lett.} {\bf 2007}, {\em 98},~230501.

\bibitem[Gisin \em{et~al.}(2010)Gisin, Pironio, and Sangouard]{Gisin2010}
Gisin, N.; Pironio, S.; Sangouard, N.
\newblock Proposal for implementing device-independent quantum key distribution
  based on a heralded qubit amplifier.
\newblock {\em Phys. Rev. Lett.} {\bf 2010}, {\em 105},~070501.

\bibitem[Vazirani and Vidick(2014)]{Vazirani2014}
Vazirani, U.; Vidick, T.
\newblock Fully device-independent quantum key distribution.
\newblock {\em Phys. Rev. Lett.} {\bf 2014}, {\em 113},~140501.

\bibitem[Grangier \em{et~al.}(1988)Grangier, Potasek, and Yurke]{Grangier1988}
Grangier, P.; Potasek, M.; Yurke, B.
\newblock Probing the phase coherence of parametrically generated photon pairs:
  A new test of Bell's inequalities.
\newblock {\em Phys. Rev. A} {\bf 1988}, {\em 38},~3132.

\bibitem[Banaszek and W{\'o}dkiewicz(1998)]{Banaszek1998}
Banaszek, K.; W{\'o}dkiewicz, K.
\newblock Nonlocality of the Einstein--Podolsky--Rosen state in the Wigner
  representation.
\newblock {\em Phys.~Rev.~A} {\bf 1998}, {\em 58},~4345--4347.

\bibitem[Banaszek and W{\'o}dkiewicz(1999)]{Banaszek1999}
Banaszek, K.; W{\'o}dkiewicz, K.
\newblock Testing quantum nonlocality in phase space.
\newblock {\em Phys. Rev. Lett.} {\bf 1999}, {\em 82},~2009--2013.

\bibitem[Marshall and Weedbrook(2014)]{Marshall2014}
Marshall, K.; Weedbrook, C.
\newblock Device-independent quantum cryptography for continuous variables.
\newblock {\em Phys. Rev. A} {\bf 2014}, {\em 90},~042311.

\bibitem[Gottesman \em{et~al.}(2001)Gottesman, Kitaev, and
  Preskill]{Gottesman2001a}
Gottesman, D.; Kitaev, A.; Preskill, J.
\newblock Encoding a qubit in an oscillator.
\newblock {\em Phys. Rev. A} {\bf 2001}, {\em 64},~012310.

\bibitem[Braunstein and Pirandola(2012)]{Braunstein2012}
Braunstein, S.L.; Pirandola, S.
\newblock Side-channel-free quantum key distribution.
\newblock {\em Phys. Rev. Lett.} {\bf 2012}, {\em 108},~130502.

\bibitem[Lo \em{et~al.}(2012)Lo, Curty, and Qi]{Lo2012}
Lo, H.K.; Curty, M.; Qi, B.
\newblock Measurement-device-independent quantum key distribution.
\newblock {\em Phys. Rev. Lett.} {\bf 2012}, {\em 108},~130503.

\bibitem[Pirandola \em{et~al.}(2015)Pirandola, Ottaviani, Spedalieri,
  Weedbrook, Braunstein, Lloyd, Gehring, Jacobsen, and Andersen]{Pirandola2014arxiv}
Pirandola, S.; Ottaviani, C.; Spedalieri, G.; Weedbrook, C.; Braunstein, S.L.;
  Lloyd, S.; Gehring, T.; Jacobsen, C.S.; Andersen, U.L.
\newblock High-rate quantum cryptography in untrusted networks. {\bf 2014}, arXiv:1312.4104. 

\bibitem[Pirandola \em{et~al.}(2015)Pirandola, Ottaviani, Spedalieri,
  Weedbrook, Braunstein, Lloyd, Gehring, Jacobsen, and Andersen]{Pirandola2015}
Pirandola, S.; Ottaviani, C.; Spedalieri, G.; Weedbrook, C.; Braunstein, S.L.;
  Lloyd, S.; Gehring, T.; Jacobsen, C.S.; Andersen, U.L.
\newblock High-rate measurement-device-independent quantum cryptography. 
\newblock {\em Nat.~Photonics} {\bf 2015}, {\em 9},~397--402.

\bibitem[Zhang \em{et~al.}(2014)Zhang, Li, Yu, Gu, Peng, and Guo]{Zhang2014}
Zhang, Y.C.; Li, Z.; Yu, S.; Gu, W.; Peng, X.; Guo, H.
\newblock Continuous-variable measurement-device-independent quantum key
  distribution using squeezed states.
\newblock {\em Phys. Rev. A} {\bf 2014}, {\em 90},~052325.

\bibitem[Li \em{et~al.}(2014)Li, Zhang, Xu, Peng, and Guo]{Li2014}
Li, Z.; Zhang, Y.C.; Xu, F.; Peng, X.; Guo, H.
\newblock Continuous-variable measurement-device-independent quantum key
  distribution.
\newblock {\em Phys. Rev. A} {\bf 2014}, {\em 89},~052301.

\bibitem[Pirandola(2014)]{Pirandola2014}
Pirandola, S.
\newblock Quantum discord as a resource for quantum cryptography.
\newblock {\em Sci. Rep.} {\bf 2014}, {\em 4}, doi:10.1038/srep06956.

\bibitem[Csisz{\'a}r and K{\"o}rner(1978)]{Csiszar1978}
Csisz{\'a}r, I.; K{\"o}rner, J.
\newblock Broadcast channels with confidential messages.
\newblock {\em IEEE Trans. Inf. Theory} {\bf 1978}, {\em
  24},~339--348.

\bibitem[Holevo and Werner(2001)]{Holevo2001}
Holevo, A.S.; Werner, R.F.
\newblock Evaluating capacities of bosonic {Gauss}ian channels.
\newblock {\em Phys. Rev. A} {\bf 2001}, {\em 63},~032312.

\bibitem[Serafini \em{et~al.}(2005)Serafini, Paris, Illuminati, and
  De~Siena]{Serafini2005}
Serafini, A.; Paris, M.G.A.; Illuminati, F.; de~Siena, S.
\newblock Quantifying decoherence in continuous variable systems.
\newblock {\em J. Opt. B: Quantum Semiclass. Opt.} {\bf 2005}, {\em 7},~doi:10.1088/1464-4266/7/4/R01.

\bibitem[Penrose(1955)]{Penrose1955}
Penrose, R.
\newblock A Generalized Inverse for Matrices.
\newblock  In \emph{Mathematical Proceedings of the Cambridge Philosophical Society};
  Cambridge University Press: Cambridge, UK, 1955; pp. 406--413.

\bibitem[Serafini \em{et~al.}(2004)Serafini, Illuminati, and
  De~Siena]{Serafini2004}
Serafini, A.; Illuminati, F.; de~Siena, S.
\newblock Symplectic invariants, entropic measures and correlations of Gaussian
  states.
\newblock {\em J. Phys. B}
  {\bf 2004}, {\em 37},~L21--L28.

\bibitem[Araki and Lieb(2002)]{Araki2002}
Araki, H.; Lieb, E.H.
\newblock Entropy Inequalities. In {\em Inequalities}; Springer: Berlin/Heidelberg, Germany, 2002; pp. 47--57.

\bibitem[Lance \em{et~al.}(2005)Lance, Symul, Sharma, Weedbrook, Ralph, and
  Lam]{Lance2005}
Lance, A.M.; Symul, T.; Sharma, V.; Weedbrook, C.; Ralph, T.C.; Lam, P.K.
\newblock No-switching quantum key distribution using broadband modulated
  coherent light.
\newblock {\em Phys. Rev. Lett.} {\bf 2005}, {\em 95},~180503.

\bibitem[Usenko \em{et~al.}(2012)Usenko, Heim, Peuntinger, Wittmann, Marquardt,
  Leuchs, and Filip]{Usenko2012}
Usenko, V.C.; Heim, B.; Peuntinger, C.; Wittmann, C.; Marquardt, C.; Leuchs,
  G.; Filip, R.
\newblock Entanglement of Gaussian states and the applicability to quantum key
  distribution over fading channels.
\newblock {\em New J. Phys.} {\bf 2012}, {\em 14},~093048.

\bibitem[Garc{\'\i}a-Patr{\'o}n and Cerf(2009)]{Garcia2009}
Garc{\'\i}a-Patr{\'o}n, R.; Cerf, N.J.
\newblock Continuous-variable quantum key distribution protocols over noisy
  channels.
\newblock {\em Phys.~Rev.~Lett.} {\bf 2009}, {\em 102},~130501.

\bibitem[Fossier \em{et~al.}(2009)Fossier, Diamanti, Debuisschert,
  Tualle-Brouri, and Grangier]{Fossier2009a}
Fossier, S.; Diamanti, E.; Debuisschert, T.; Tualle-Brouri, R.; Grangier, P.
\newblock Improvement of continuous-variable quantum key distribution systems
  by using optical preamplifiers.
\newblock {\em J. Phys. B}
  {\bf 2009}, {\em 42},~114014.

\bibitem[Filip(2008)]{Filip2008}
Filip, R.
\newblock Continuous-variable quantum key distribution with noisy coherent
  states.
\newblock {\em Phys. Rev. A} {\bf 2008}, {\em 77},~022310.

\bibitem[Usenko and Filip(2010)]{Usenko2010a}
Usenko, V.C.; Filip, R.
\newblock Feasibility of continuous-variable quantum key distribution with
  noisy coherent states.
\newblock {\em Phys.~Rev.~A} {\bf 2010}, {\em 81},~022318.

\bibitem[Weedbrook \em{et~al.}(2010)Weedbrook, Pirandola, Lloyd, and
  Ralph]{Weedbrook2010}
Weedbrook, C.; Pirandola, S.; Lloyd, S.; Ralph, T.C.
\newblock Quantum cryptography approaching the classical limit.
\newblock {\em Phys.~Rev.~Lett.} {\bf 2010}, {\em 105},~110501.

\bibitem[Weedbrook \em{et~al.}(2012)Weedbrook, Pirandola, and
  Ralph]{Weedbrook2012}
Weedbrook, C.; Pirandola, S.; Ralph, T.C.
\newblock Continuous-variable quantum key distribution using thermal states.
\newblock {\em Phys.~Rev.~A} {\bf 2012}, {\em 86},~022318.

\bibitem[Jouguet \em{et~al.}(2012)Jouguet, Kunz-Jacques, Diamanti, and
  Leverrier]{Jouguet2012}
Jouguet, P.; Kunz-Jacques, S.; Diamanti, E.; Leverrier, A.
\newblock Analysis of imperfections in practical continuous-variable quantum
  key distribution.
\newblock {\em Phys. Rev. A} {\bf 2012}, {\em 86},~032309.

\bibitem[Huang \em{et~al.}(2013)Huang, Weedbrook, Yin, Wang, Li, Chen, Guo, and
  Han]{Huang2013}
Huang, J.Z.; Weedbrook, C.; Yin, Z.Q.; Wang, S.; Li, H.W.; Chen, W.; Guo, G.C.;
  Han, Z.F.
\newblock Quantum hacking of a continuous-variable quantum-key-distribution
  system using a wavelength attack.
\newblock {\em Phys. Rev. A} {\bf 2013}, {\em 87},~062329.

\bibitem[Ma \em{et~al.}(2013)Ma, Sun, Jiang, and Liang]{Ma2013}
Ma, X.C.; Sun, S.H.; Jiang, M.S.; Liang, L.M.
\newblock Wavelength attack on practical continuous-variable
  quantum-key-distribution system with a heterodyne protocol.
\newblock {\em Phys. Rev. A} {\bf 2013}, {\em 87},~052309.

\bibitem[Huang \em{et~al.}(2014)Huang, Kunz-Jacques, Jouguet, Weedbrook, Yin,
  Wang, Chen, Guo, and Han]{Huang2014}
Huang, J.Z.; Kunz-Jacques, S.; Jouguet, P.; Weedbrook, C.; Yin, Z.Q.; Wang, S.;
  Chen, W.; Guo, G.C.; Han, Z.F.
\newblock Quantum hacking on quantum key distribution using homodyne detection.
\newblock {\em Phys. Rev. A} {\bf 2014}, {\em 89},~032304.

\bibitem[Kunz-Jacques and Jouguet(2015)]{Kunz2015}
Kunz-Jacques, S.; Jouguet, P.
\newblock Robust shot-noise measurement for continuous-variable quantum key
  distribution.
\newblock {\em Phys.~Rev. A} {\bf 2015}, {\em 91},~022307.

\bibitem[Jouguet \em{et~al.}(2013)Jouguet, Kunz-Jacques, and
  Diamanti]{Jouguet2013a}
Jouguet, P.; Kunz-Jacques, S.; Diamanti, E.
\newblock Preventing calibration attacks on the local oscillator in
  continuous-variable quantum key distribution.
\newblock {\em Phys. Rev. A} {\bf 2013}, {\em 87},~062313.

\bibitem[Ma \em{et~al.}(2013)Ma, Sun, Jiang, and Liang]{Ma2013a}
Ma, X.C.; Sun, S.H.; Jiang, M.S.; Liang, L.M.
\newblock Local oscillator fluctuation opens a loophole for Eve in practical
  continuous-variable quantum-key-distribution systems.
\newblock {\em Phys. Rev. A} {\bf 2013}, {\em 88},~022339.

\bibitem[Ma \em{et~al.}(2014)Ma, Sun, Jiang, Gui, Zhou, and Liang]{Ma2014}
Ma, X.C.; Sun, S.H.; Jiang, M.S.; Gui, M.; Zhou, Y.L.; Liang, L.M.
\newblock Enhancement of the security of a practical continuous-variable
  quantum-key-distribution system by manipulating the intensity of the local
  oscillator.
\newblock {\em Phys.~Rev.~A} {\bf 2014}, {\em 89},~032310.

\bibitem[Shen \em{et~al.}(2011)Shen, Peng, Yang, and Guo]{Shen2011}
Shen, Y.; Peng, X.; Yang, J.; Guo, H.
\newblock Continuous-variable quantum key distribution with Gaussian source
  noise.
\newblock {\em Phys. Rev. A} {\bf 2011}, {\em 83},~052304.

\bibitem[Huang \em{et~al.}(2013)Huang, He, and Zeng]{Huang2013a}
Huang, P.; He, G.Q.; Zeng, G.H.
\newblock Bound on Noise of Coherent Source for Secure Continuous-Variable
  Quantum Key Distribution.
\newblock {\em Int. J. Theor. Phys.} {\bf 2013}, {\em
  52},~1572--1582.

\bibitem[Yang \em{et~al.}(2012)Yang, Xu, and Guo]{Yang2012}
Yang, J.; Xu, B.; Guo, H.
\newblock Source monitoring for continuous-variable quantum key distribution.
\newblock {\em Phys. Rev. A} {\bf 2012}, {\em 86},~042314.

\bibitem[Wang \em{et~al.}(2014)Wang, Yu, Zhang, Gu, and Guo]{Wang2014a}
Wang, T.; Yu, S.; Zhang, Y.C.; Gu, W.; Guo, H.
\newblock Improving the maximum transmission distance of continuous-variable
  quantum key distribution with noisy coherent states using a noiseless
  amplifier.
\newblock {\em Phys.~Lett.~A} {\bf 2014}, {\em 378},~2808--2812.

\bibitem[Madsen \em{et~al.}(2012)Madsen, Usenko, Lassen, Filip, and
  Andersen]{Madsen2012}
Madsen, L.S.; Usenko, V.C.; Lassen, M.; Filip, R.; Andersen, U.L.
\newblock Continuous variable quantum key distribution with modulated entangled
  states.
\newblock {\em Nat. Commun.} {\bf 2012}, {\em 3}, doi:10.1038/ncomms2097.

\bibitem[Weedbrook(2013)]{Weedbrook2013}
Weedbrook, C.
\newblock Continuous-variable quantum key distribution with entanglement in the
  middle.
\newblock {\em Phys. Rev. A} {\bf 2013}, {\em 87},~022308.

\bibitem[Su \em{et~al.}(2009)Su, Wang, Wang, Jia, Xie, and Peng]{Su2009}
Su, X.; Wang, W.; Wang, Y.; Jia, X.; Xie, C.; Peng, K.
\newblock Continuous variable quantum key distribution based on optical
  entangled states without signal modulation.
\newblock {\em Europhys. Lett.} {\bf 2009}, {\em 87},~20005.

\bibitem[Usenko \em{et~al.}(2014)Usenko, Ruppert, and Filip]{Usenko2014}
Usenko, V.C.; Ruppert, L.; Filip, R.
\newblock Entanglement-based continuous-variable quantum key distribution with
  multimode states and detectors.
\newblock {\em Phys. Rev. A} {\bf 2014}, {\em 90},~062326.

\bibitem[Usenko \em{et~al.}(2015)Usenko, Ruppert, and Filip]{Usenko2015a}
Usenko, V.C.; Ruppert, L.; Filip, R.
\newblock Quantum communication with macroscopically bright nonclassical
  states.
\newblock {\em Opt.~Express} {\bf 2015}, {\em 23},~31534--31543.

\bibitem[Pe{\v r}ina \em{et~al.}(2007)Pe{\v r}ina, K{\v r}epelka, Pe{\v r}ina,
  Bondani, Allevi, and Andreoni]{Perina2007}
Pe{\v r}ina, J.; K{\v r}epelka, J.; Pe{\v r}ina, J.; Bondani, M.; Allevi, A.;
  Andreoni, A.
\newblock Experimental joint signal-idler quasidistributions and photon-number
  statistics for mesoscopic twin beams.
\newblock {\em Phys. Rev. A} {\bf 2007}, {\em 76},~043806.

\bibitem[Iskhakov \em{et~al.}(2009)Iskhakov, Chekhova, and
  Leuchs]{Iskhakov2009}
Iskhakov, T.; Chekhova, M.V.; Leuchs, G.
\newblock Generation and direct detection of broadband mesoscopic
  polarization-squeezed vacuum.
\newblock {\em Phys. Rev. Lett.} {\bf 2009}, {\em 102},~183602.

\bibitem[Pirandola \em{et~al.}(2008)Pirandola, Mancini, Lloyd, and
  Braunstein]{Pirandola2008a}
Pirandola, S.; Mancini, S.; Lloyd, S.; Braunstein, S.L.
\newblock Continuous-variable quantum cryptography using two-way quantum
  communication.
\newblock {\em Nat. Phys.} {\bf 2008}, {\em 4},~726--730.

\bibitem[Weedbrook \em{et~al.}(2014)Weedbrook, Ottaviani, and
  Pirandola]{Weedbrook2014}
Weedbrook, C.; Ottaviani, C.; Pirandola, S.
\newblock Two-way quantum cryptography at different wavelengths.
\newblock {\em Phys. Rev. A} {\bf 2014}, {\em 89},~012309.

\bibitem[Wang \em{et~al.}(2014)Wang, Yu, Zhang, Gu, and Guo]{Wang2014}
Wang, T.; Yu, S.; Zhang, Y.C.; Gu, W.; Guo, H.
\newblock Security of two-way continuous-variable quantum key distribution with
  source noise.
\newblock {\em J. Phys. B}
  {\bf 2014}, {\em 47},~215504.

\bibitem[da~Silva \em{et~al.}(2010)da~Silva, Bozyigit, Wallraff, and
  Blais]{Silva2010}
Da~Silva, M.P.; Bozyigit, D.; Wallraff, A.; Blais, A.
\newblock Schemes for the observation of photon correlation functions in
  circuit QED with linear detectors.
\newblock {\em Phys. Rev. A} {\bf 2010}, {\em 82},~043804.

\bibitem[Bozyigit \em{et~al.}(2011)Bozyigit, Lang, Steffen, Fink, Eichler,
  Baur, Bianchetti, Leek, Filipp, Da~Silva, et~al.]{Bozyigit2011}
Bozyigit, D.; Lang, C.; Steffen, L.; Fink, J.; Eichler, C.; Baur, M.;
  Bianchetti, R.; Leek, P.; Filipp, S.; da~Silva,~M.;~\emph{et~al}.~\newblock Antibunching of microwave-frequency photons observed in correlation
  measurements using linear detectors.
\newblock {\em Nat. Phys.} {\bf 2011}, {\em 7}, 154--158.

\bibitem[Eichler \em{et~al.}(2011)Eichler, Bozyigit, Lang, Baur, Steffen, Fink,
  Filipp, and Wallraff]{Eichler2011}
Eichler, C.; Bozyigit, D.; Lang, C.; Baur, M.; Steffen, L.; Fink, J.; Filipp,
  S.; Wallraff, A.
\newblock Observation of two-mode squeezing in the microwave frequency domain.
\newblock {\em Phys. Rev. Lett.} {\bf 2011}, {\em 107},~113601.

\end{thebibliography}


%


%

\end{document}